\documentclass[english]{IEEEtran}
\usepackage[T1]{fontenc}
\usepackage[latin9]{inputenc}
\usepackage{float}
\usepackage{calc}
\usepackage{amsmath}
\usepackage{amsthm}
\usepackage{amssymb}
\usepackage{graphicx}

\makeatletter

\providecommand{\tabularnewline}{\\}
\floatstyle{ruled}
\newfloat{algorithm}{tbp}{loa}
\providecommand{\algorithmname}{Algorithm}
\floatname{algorithm}{\protect\algorithmname}

\theoremstyle{definition}
\newtheorem{example}{\protect\examplename}
\theoremstyle{plain}
\newtheorem{prop}{\protect\propositionname}
\theoremstyle{remark}
\newtheorem{rem}{\protect\remarkname}
\theoremstyle{plain}
\newtheorem{lem}{\protect\lemmaname}
\theoremstyle{plain}
\newtheorem{thm}{\protect\theoremname}

\usepackage{cite}

\newtheorem{assumption}{Assumption}
\newtheorem{challenge}{Challenge}

\makeatother

\usepackage{babel}
\providecommand{\examplename}{Example}
\providecommand{\lemmaname}{Lemma}
\providecommand{\propositionname}{Proposition}
\providecommand{\remarkname}{Remark}
\providecommand{\theoremname}{Theorem}

\begin{document}

\title{Stochastic Successive Convex Approximation for Non-Convex Constrained
Stochastic Optimization }

\author{{\normalsize{}An Liu$^{1}$, }\textit{\normalsize{}Senior Member,
IEEE}{\normalsize{}, Vincent Lau$^{2}$,}\textit{\normalsize{} Fellow
IEEE}{\normalsize{} and Borna Kananian$^{3}$, }\textit{\normalsize{}Student
Member, IEEE}{\normalsize{}\\$^{1}$College of Information Science
and Electronic Engineering, Zhejiang University\\$^{2}$Department
of Electronic and Computer Engineering, Hong Kong University of Science
and Technology\\$^{3}$Department of Electrical Engineering, Sharif
University of Technology}\thanks{This work was supported by the National Science Foundation of China
under Project No. 61571383 and RGC 16209916.}}
\maketitle
\begin{abstract}
This paper proposes a constrained stochastic successive convex approximation
(CSSCA) algorithm to find a stationary point for a general non-convex
stochastic optimization problem, whose objective and constraint functions
are non-convex and involve expectations over random states. Most existing
methods for non-convex stochastic optimization, such as the stochastic
(average) gradient and stochastic majorization-minimization, only
consider minimizing a stochastic non-convex objective over a deterministic
convex set. The proposed CSSCA algorithm can also handle stochastic
non-convex constraints in optimization problems, and it opens the
way to solving more challenging optimization problems that occur in
many applications. The algorithm is based on solving a sequence of
convex objective/feasibility optimization problems obtained by replacing
the objective/constraint functions in the original problems with some
convex surrogate functions. The CSSCA algorithm allows a wide class
of surrogate functions and thus provides many freedoms to design good
surrogate functions for specific applications. Moreover, it also facilitates
parallel implementation for solving large scale stochastic optimization
problems, which arise naturally in today's signal processing such
as machine learning and big data analysis. We establish the convergence
of CSSCA algorithm with a feasible initial point, and customize the
algorithmic framework to solve several important application problems.
Simulations show that the CSSCA algorithm can achieve superior performance
over existing solutions.
\end{abstract}


\section{Introduction}

\subsection{Background}

Deterministic convex optimization theory is very powerful and allows
low complexity solutions for large scale problems. However, stochastic
processes and effects appear naturally in the real physical world
and in many cases, their effects cannot be neglected. For example,
in wireless communications, we have random channel fading as well
as random noise and interference at the receiver. In signal processing
applications, such as radar detection or signal recovery, we need
to extract useful signals and data from those that are contaminated
in noisy observations. In all these examples, the physical system
is not deterministic and it is naturally important to take into account
the underlying random process in modeling optimization problems. This
motivates the study of stochastic optimization. In fact, stochastic
optimizations play a critical role in various key application areas
such as wireless resource optimizations, compressive sensing and (sparse)
signal recovery, machine learning, etc.

Despite the important role of stochastic optimization in many applications,
it is still far from mature compared to its deterministic counterpart.
For example, we still lack an efficient algorithm to solve non-convex
stochastic optimization problems that occur in many applications,
especially when the constraint is also non-convex and involves expectations
over random states. Moreover, many applications dealing with large
systems require solving large scale (non-convex) stochastic optimization
problems. In this case, it is desirable to design parallel algorithms
that can distribute the computational load across a number of computation
nodes. In this paper, we propose a constrained stochastic successive
convex approximation (CSSCA) method for general non-convex stochastic
optimization problems whose objective and constraints contain expectations
of non-convex functions. The CSSCA method is also suitable for parallel
implementation.

\subsection{Related works}

There are three major existing methods on non-convex stochastic optimization.

\textbf{Stochastic Gradient-based methods:} Stochastic gradient/subgradient
\cite{Spall_Wiley03_SO} is a common method to solve unconstrained
stochastic optimization problems. In each iteration, an unbiased estimation
of the gradient of the objective function is obtained and a gradient-like
update is performed. Under some technical conditions, almost sure
convergence to stationary points can be established \cite{Bertsekas_SIAM2000_SG}.
Various variations of the stochastic gradient method have been proposed
\cite{Polyak_SIAM1992_SO,Ram_Asilomar07_ISG,NIPS2013_4937,NIPS2014_5258}.
For convex stochastic optimization problems with a simple convex feasible
set, the stochastic gradient projection method has been proposed and
been shown to converge to the optimal solution almost surely \cite{Ermoliev_Cybern1972_SG,Yousefian_Automatica2012_SG}.
To better handle the non-convexity, a gradient averaging method \cite{Ruszczyski_MP1980_SAG,Bach_2014AAS_SAG}
is proposed where the gradient projection update at each iteration
is based on the average of the current and past gradient samples.
Intuitively, the average sample gradient tends to converge to the
true gradient of the objective function and thus the convergence follows
a similar analysis to that of the gradient projection method for deterministic
non-convex problems. Under some technical conditions, one can indeed
prove the convergence of the gradient averaging method to a stationary
point \cite{Gupal_Cyber1972_SAG}. Algorithms with averaging in both
gradients and iterates (optimization variables) are proposed, where
at each iteration, an average gradient is used for the gradient projection
update, and the output is also given by the average of the current
and past iterates \cite{Yin_SSR_1994_SGA,Yin1995}.

\textbf{Stochastic Majorization-Minimization: }Majorization-minimization
(MM) \cite{Sun_TSP2017_MM} is a powerful optimization principle that
includes many well-known optimization methods as special cases, such
as proximal gradient method \cite{Bertsekas_TR2010_proximal}, expectation-maximization
(EM) algorithm \cite{Olivier_JRS2009_EM}, cyclic minimization \cite{Stoica_spm2004_CyclicMIN},
and variational Bayes techniques \cite{Wainwright_FTML2008_VBI}.
The basic idea of MM is to iteratively minimize a surrogate function
that upper-bounds the objective (but matches the value of the objective
function and its derivative at the current iterate). MM monotonically
decreases the objective value until convergent to a stationary point.
Stochastic MM \cite{NIPS2013_5129_StochasticMM,Chouzenoux_TSP17_StochasticMM}
is an extension of MM to solve stochastic non-convex optimization
problems. Specifically, at each iteration, a \textit{sample surrogate
function} is first obtained as an upper bound of the sample objective
function. Then the updated optimization variable is obtained by minimizing
the average surrogate function (the average of the current and past
sample surrogate functions). Intuitively, the average surrogate function
tends to converge to a deterministic upper bound of the objective
function that matches the value of the objective function and its
derivative at a limiting point, from which it can be shown that any
limiting point of the algorithm is a stationary point. Please refer
to \cite{NIPS2013_5129_StochasticMM,Chouzenoux_TSP17_StochasticMM}
for the formal convergence proof of the stochastic MM.

\textbf{Stochastic Successive Convex Approximation (SCA): }SCA \cite{Scutari_TSP14_SCA}
is similar to MM in the sense that it also iteratively minimizes a
sequence of surrogate functions. However, the conditions on the surrogate
functions are different. SCA requires the surrogate function to be
convex but not necessarily an upper bound of the objective function.
On the other hand, MM requires the surrogate function to be an upper
bound of the objective function but not necessarily convex\footnote{In practice, the surrogate function used in MM is usually convex for
complexity consideration.}. Since there is no upper bound constraint, we have more freedom to
choose a surrogate function at each iteration that can better approximate
the objective function. As a result, SCA may yield a faster convergence
speed with properly chosen surrogate functions. In \cite{Yang_TSP2016_SSCA},
a stochastic parallel SCA method is proposed for non-convex stochastic
sum-utility optimization problems in multi-agent networks. In this
method, all agents update their optimization variables in parallel
by solving a sequence of convex subproblems. Almost sure convergence
to stationary points is also proved.

\subsection{Contributions}

All of the above existing works on non-convex stochastic optimization
have assumed simple constraints where the feasible set of the problem
can be represented by a deterministic convex set. However, in many
applications, such as those considered in Section \ref{sec:System-Model},
the constraints may involve expectations of non-convex functions.
Moreover, there are few works on parallel algorithms that are suitable
for large scale non-convex stochastic optimization, and the existing
parallel algorithms such as the parallel SCA method in \cite{Yang_TSP2016_SSCA}
often assume that the constraint can be represented by a Cartesian
product of deterministic convex sets, which significantly limits their
applications. In this paper, we propose a more general non-convex
stochastic optimization method to avoid many of the above restrictions
on the objective/constraints. The main contributions are summarized
below.
\begin{itemize}
\item \textbf{A general stochastic SCA method and its convergence proof:}
We propose a CSSCA method which can be applied to more general non-convex
stochastic optimization problems whose objective and constraint contain
expectations of non-convex functions. This opens the door for solving
more difficult stochastic optimization problems that occur in many
new applications. Moreover, we establish the convergence of CSSCA
method to stationary points for the case when the initial point is
feasible. Specifically, based on the asymptotic consistency (i.e.,
the values and gradients of surrogate functions asymptotically match
the original objective/constraint functions at the current iterate)
and strong convexity assumption of surrogate functions, we first use
contradiction to show that all limiting points must be feasible w.p.1.
Then we show that every limiting point must be a stationary point
of the convex optimization subproblem associated with the surrogate
functions, from which and the asymptotic consistency of surrogate
functions, it can be shown that any limiting point of the algorithm
is also a stationary point of the original problem w.p.1.  
\item \textbf{Parallel CSSCA:} We propose a parallel CSSCA algorithm where
the minimization of the surrogate function is decomposed into independent
subproblems and each subproblem is solved by a user (computation node)
in a parallel way. Such a parallel CSSCA algorithm is suitable for
solving large-scale (non-convex) stochastic optimization problems
arising in machine learning and signal processing.
\item \textbf{Specific CSSCA algorithm design for some important applications:}
We apply the CSSCA to solve several important application problems
in wireless communications. We show that it is crucial to choose application
specific surrogate functions for different applications. We believe
that the proposed CSSCA-based solutions for these application problems
alone are of great interest to the community.
\end{itemize}

The rest of the paper is organized as follows. The problem formulation
is given in Section \ref{sec:System-Model}, together with some application
examples. The CSSCA algorithm and the convergence analysis are presented
in Section \ref{sec:Constrained-Stochastic-Successiv} and \ref{sec:Convergence-Analysis},
respectively. The parallel CSSCA algorithm is proposed in Section
\ref{sec:Parallel-Implementation-for}. Section \ref{sec:Applications}
applies the CSSCA method to solve several important application problems.
Finally, the conclusion is given in Section \ref{sec:Conclusion}.

\section{Problem Formulations\label{sec:System-Model}}

Consider the following non-convex constrained stochastic optimization
problem:
\begin{align}
\min_{\boldsymbol{x}\in\mathcal{X}} & f_{0}\left(\boldsymbol{x}\right)\triangleq\mathbb{E}\left[g_{0}\left(\boldsymbol{x},\xi\right)\right]\label{eq:mainP}\\
s.t. & f_{i}\left(\boldsymbol{x}\right)\triangleq\mathbb{E}\left[g_{i}\left(\boldsymbol{x},\xi\right)\right]\leq0,i=1,....,m,\nonumber 
\end{align}
where $\boldsymbol{x}\in\mathcal{X}$ is the optimization variable
with $\mathcal{X}$ being the domain of the problem; and $\boldsymbol{\xi}$
is a random state defined on the probability space $\left(\Omega,\mathcal{F},\mathbb{P}\right)$,
with $\Omega$ being the sample space, $\mathcal{F}$ being the $\sigma$-algebra
generated by subsets of $\Omega$, and $\mathbb{P}$ being a probability
measure defined on $\mathcal{F}$. We make the following assumptions
on the problem structure.

\begin{assumption}[Assumptions on Problem (\ref{eq:mainP})]\label{asm:convP}$\:$
\begin{enumerate}
\item $\mathcal{X}\subseteq\mathbb{R}^{n_{x}}$ for some positive integer
$n_{x}$. Moreover, $\mathcal{X}$ is compact and convex.
\item The functions $g_{i}:\text{ }\mathcal{X}\times\Omega\mapsto\mathbb{R},i=0,...,m$
are continuously differentiable (and possibly non-convex) functions
in $\boldsymbol{x}$.
\item For any $i\in\left\{ 0,...,m\right\} $ and $\boldsymbol{\xi}\in\Omega$,
the function $g_{i}\left(\boldsymbol{x},\boldsymbol{\xi}\right)$,
its derivative, and its second order derivative are uniformly bounded.
\end{enumerate}
\end{assumption}

The smoothness condition in the above assumption is necessary for
both the surrogate function design and convergence proof. For example,
the construction of the two example surrogate function designs in
Section \ref{subsec:Smooth-Surrogate-Function} requires the existence
of the gradients of $g_{i}$'s. The convergence analysis in Section
\ref{sec:Convergence-Analysis} is also based on the KKT conditions
for optimization problems with smooth objective/constraint functions.
Note that although we assume $\boldsymbol{x}$ is real vectors for
clarity, the proposed algorithm can be directly applied to the case
with complex optimization variables $\boldsymbol{x}$, by treating
each function $g_{i}\left(\boldsymbol{x},\xi\right)$ in the problem
as a real valued function of real vectors $\left[\textrm{Re}\left[\boldsymbol{x}\right];\textrm{Im}\left[\boldsymbol{x}\right]\right]$.
Problem (\ref{eq:mainP}) embraces a lot of important applications
including chance constraint problems \cite{Nemirovski_SIAM2006_CCp}.
In the following, we give some important application examples of the
problem formulation in (\ref{eq:mainP}).
\begin{example}
[MIMO Transmit Signal Design with Imperfect CSI \cite{Ding_TSP09_MIMOimpCSI}]\label{exa:Robust-MIMO-Transmit}Consider
a downlink system that consists of a multiple-antenna base station
(BS) and $K$ single-antenna users. The BS is equipped with $n$ antennas,
and it simultaneously transmits $K$ data streams to the $K$ users
using MIMO signaling based on the estimated channel state information
(CSI) $\hat{\boldsymbol{h}}_{k},k=1,...,K$. The true channel vectors
$\boldsymbol{h}_{k}$'s can be modeled as $\boldsymbol{h}_{k}=\hat{\boldsymbol{h}}_{k}+\boldsymbol{e}_{k}$,
where $\boldsymbol{e}_{k}$ represents the channel estimation error.
With channel estimation error, the BS can no longer guarantee the
desired rate for each user. In this case, the BS may improve the average
MIMO transmission performance under the channel estimation error by
ensuring that the expected rate of each user must exceed a target
value. Specifically, the MIMO transmit signal design problem with
imperfect CSI can be formulated as the following power minimization
problem subject to the expected rate requirement:
\begin{align}
\min_{\left\{ \boldsymbol{Q}_{k}\succeq\boldsymbol{0}\right\} } & \sum_{k=1}^{K}Tr\left(\boldsymbol{Q}_{k}\right)\label{eq:RBF}\\
s.t. & \mathbb{E}\left[\log\left(1+\frac{\boldsymbol{h}_{k}^{H}\boldsymbol{Q}_{k}\boldsymbol{h}_{k}}{\sum_{j\neq k}\boldsymbol{h}_{k}^{H}\boldsymbol{Q}_{j}\boldsymbol{h}_{k}+\sigma_{k}^{2}}\right)\right]\geq r_{k},\forall k,\nonumber 
\end{align}
where $\boldsymbol{Q}_{k}$ is the covariance matrix of the transmit
signal for user $k$, $\sigma_{k}^{2}$ is the variance of the thermal
noise at user $k$, and $r_{k}$ is the expected rate requirement
for user $k$. The expectation is taken w.r.t. the channel estimation
error $\boldsymbol{e}_{k}$ conditioned on $\hat{\boldsymbol{h}}_{k}$.
In Problem (\ref{eq:RBF}), the random state is $\boldsymbol{\xi}=\left[\boldsymbol{e}_{1},...,\boldsymbol{e}_{K}\right]^{T}$.
The \textit{sample objective function} $g_{0}\left(\boldsymbol{x},\xi\right)$
is convex, and the \textit{sample constraint functions} $g_{i}\left(\boldsymbol{x},\xi\right),i=1,...,K$
are non-convex.
\end{example}

\begin{example}
[Robust Beamforming Design \cite{Wang_ICASSP11_Bernstein-type}]\label{exa:Portfolio-optimization}Consider
the same MIMO downlink system with channel estimation error as in
Example \ref{exa:Robust-MIMO-Transmit}. However, unlike Example \ref{exa:Robust-MIMO-Transmit}
where the expected rate of each user is guaranteed under the channel
estimation error, we consider a stronger quality of service requirement
where the rate of each user must exceed a target value with high probability.
To be more specific, we consider the following robust beamforming
design formulation:
\begin{align}
\min_{\left\{ \boldsymbol{w}_{k}\right\} } & \sum_{k=1}^{K}\left\Vert \boldsymbol{w}_{k}\right\Vert ^{2}\nonumber \\
s.t. & \Pr\left[SINR_{k}\triangleq\frac{\left|\boldsymbol{h}_{k}^{H}\boldsymbol{w}_{k}\right|^{2}}{\sum_{i\neq k}\left|\boldsymbol{h}_{k}^{H}\boldsymbol{w}_{i}\right|^{2}+\sigma_{k}^{2}}\leq\eta_{k}\right]\leq\epsilon,\label{eq:RBFCcin}
\end{align}
where $\boldsymbol{w}_{k}\in\mathbb{C}^{n}$ is the beamforming vector
for user $k$, $\sum_{k=1}^{K}\left\Vert \boldsymbol{w}_{k}\right\Vert ^{2}$
is the total transmit power at the BS, and the constraint (\ref{eq:RBFCcin})
ensures that the SINR of user $k$ exceeds a target value $\eta_{k}$
with probability no less than $1-\epsilon$. Note that the BS only
knows $\hat{\boldsymbol{h}}_{k}$. Therefore, (\ref{eq:RBFCcin})
is a chance constraint with the random state given by the channel
estimation error vector $\boldsymbol{\xi}=\left[\boldsymbol{e}_{1},...,\boldsymbol{e}_{K}\right]^{T}$. 

Problem (\ref{eq:RBFCcin}) is a chance constrained problem \cite{Nemirovski_SIAM2006_CCp}
and is not exactly an instance of Problem (\ref{eq:mainP}). However,
we can transform Problem (\ref{eq:RBFCcin}) into an approximate formulation
which is an instance of Problem (\ref{eq:mainP}) as follows. First,
note that $\Pr\left[SINR_{k}\leq\eta_{k}\right]=\mathbb{E}\left[u\left(\eta_{k}-SINR_{k}\right)\right]$,
where $u\left(x\right)$ is the step function. There are many smooth
approximations of the step function. Let $\hat{u}_{\theta}\left(x\right)$
denote a smooth approximation of the step function $u\left(x\right)$
with smooth parameter $\theta$, e.g., one possible form of a smooth
approximate function is
\begin{equation}
\hat{u}_{\theta}\left(x\right)=\frac{1}{1+e^{-\theta x}},\label{eq:ustep}
\end{equation}
where the smooth parameter $\theta$ can be used to control the approximation
error. By replacing the step $u\left(x\right)$ with its smooth approximation
$\hat{u}_{\theta}\left(x\right)$, we can obtain an approximation
of Problem (\ref{eq:RBFCcin}):

\begin{align}
\min_{\left\{ \boldsymbol{w}_{k}\right\} } & \sum_{k=1}^{K}\left\Vert \boldsymbol{w}_{k}\right\Vert ^{2}\nonumber \\
s.t. & \mathbb{E}\left[\hat{u}_{\theta}\left(\eta_{k}\left(\sum_{i\neq k}\left|\boldsymbol{h}_{k}^{H}\boldsymbol{w}_{i}\right|^{2}+\sigma_{k}^{2}\right)-\left|\boldsymbol{h}_{k}^{H}\boldsymbol{w}_{k}\right|^{2}\right)\right]\leq\epsilon,\label{eq:RBFCcin-1}
\end{align}
which is an instance of Problem (\ref{eq:mainP}). Using the above
approximation, a general chance constrained problem can also be transformed
into Problem (\ref{eq:mainP}). 
\end{example}

\begin{example}
[Massive MIMO Hybrid Beamforming Design \cite{Liu_TSP14_RFprecoding}]\label{exa:Massive-MIMO-Hybrid}Consider
a multi-user massive MIMO downlink system where a BS serves $K$ single-antenna
users. The BS is equipped with $M\gg1$ antennas and $S$ transmit
RF chains, where $K\leq S<M$. Hybrid beamforming \cite{Liu_TSP14_RFprecoding,Liu_TSP2016_CSImassive}
is employed at the BS to support simultaneous transmissions to the
$K$ users. Specifically, the precoder is split into a baseband precoder
and an RF precoder as $\boldsymbol{F}\boldsymbol{G}$, where $\boldsymbol{G}=\left[\boldsymbol{g}_{1},...,\boldsymbol{g}_{K}\right]\in\mathbb{C}^{S\times K}$
is the baseband precoder using the $S$ RF chains, and $\boldsymbol{F}\in\mathbb{C}^{M\times S}$
is the RF precoder using, for example, the RF phase shifting network
\cite{Zhang_TSP05_RFshifter}. Hence, all elements of $\boldsymbol{F}$
have equal magnitude, i.e., $F_{m,s}=e^{j\theta_{m,s}}$, where $\theta_{m,s}$
is the phase of the $\left(m,s\right)$-th element $F_{m,s}$ of $\boldsymbol{F}$.
For given RF precoder $\boldsymbol{F},$ a regularized zero-forcing
(RZF) baseband precoder is used to mitigate the multi-user interference,
i.e., 
\[
\boldsymbol{G}=\boldsymbol{F}^{H}\boldsymbol{H}^{H}\left(\boldsymbol{H}\boldsymbol{F}\boldsymbol{F}^{H}\boldsymbol{H}^{H}+\frac{K}{P}\boldsymbol{I}\right)^{-1}\boldsymbol{P}^{1/2},
\]
where $\boldsymbol{H}=\left[\boldsymbol{h}_{k}\right]_{k=1,...,K}^{H}\in\mathbb{C}^{K\times M}$
is the composite channel matrix, $\boldsymbol{h}_{k}\in\mathbb{C}^{M}$
is the channel vector of user $k$, $\boldsymbol{P}=Diag\left(p_{1},...,p_{K}\right)$
with $p_{k}$ representing a parameter to control the tradeoff between
the transmit power allocated to user $k$ and the data rate of user
$k$, and $P$ is the average transmit power constraint. Consider
the maximization of the ergodic sum rate in the above massive MIMO
system with hybrid beamforming:

\begin{align}
\max_{\boldsymbol{\Theta},\boldsymbol{p}} & \sum_{k=1}^{K}\mathbb{E}\left[\log\left(1+\frac{\left|\boldsymbol{h}_{k}^{H}\boldsymbol{F}\boldsymbol{g}_{k}\right|^{2}}{\sum_{i\neq k}\left|\boldsymbol{h}_{k}^{H}\boldsymbol{F}\boldsymbol{g}_{i}\right|^{2}+1}\right)\right]\label{eq:NSP}\\
s.t. & \mathbb{E}\left[Tr\left(\boldsymbol{F}\boldsymbol{G}\boldsymbol{G}^{H}\boldsymbol{F}^{H}\right)\right]-P\leq0,\nonumber 
\end{align}
where $\boldsymbol{\Theta}\in\mathbb{C}^{M\times S}$ and the $\left(m,s\right)$-th
element of $\boldsymbol{\Theta}$ is $\theta_{m,s}$, and $\boldsymbol{p}=\left[p_{1},...,p_{K}\right]^{T}$.
Note that $\boldsymbol{F}$ is a function of $\boldsymbol{\Theta}$
and $\boldsymbol{G}$ is a function of $\boldsymbol{\Theta},\boldsymbol{p}$.
Problem (\ref{eq:NSP}) is an instance of Problem (\ref{eq:mainP})
with random state $\boldsymbol{H}$.
\end{example}
Note that in Example 1 and 2, there is no bounded constraint on $\mathcal{X}$
explicitly. In the simulations, it is observed that the iterates generated
by the algorithm is still bounded even without explicitly imposing
a bounded constraint. In practical applications, the optimization
variables are almost always bounded and we can easily add some simple
bounded constraints (such as a box region constraint) with a sufficiently
large boundary to make $\mathcal{X}$ compact, without destroying
the optimality.

\section{Constrained Stochastic Successive Convex Approximation\label{sec:Constrained-Stochastic-Successiv}}

\subsection{Challenges of Solving Problem (\ref{eq:mainP})}

Since Problem (\ref{eq:mainP}) is, in general, non-convex, we focus
on designing an efficient algorithm to find a stationary point of
Problem (\ref{eq:mainP}). There are two major challenges in solving
Problem (\ref{eq:mainP}): 1) the non-convexity of the constraint
functions; and 2) the stochastic nature of the constraint functions
(i.e., it is difficult to accurately calculate the expectations in
the constraint functions). 

For the special case when $\xi$ is a deterministic vector, (\ref{eq:mainP})
reduces to a deterministic optimization problem with non-convex constraint.
In this case, an MM algorithm has been proposed in \cite{Meisam_thesis14_BSUM}
to find a stationary point. The MM algorithm in \cite{Meisam_thesis14_BSUM}
starts from a feasible point. Due to the property of MM, it can be
shown that all the subsequent iterates generated by the MM algorithm
are still feasible, and the algorithm will eventually converge to
a stationary point. However, in the stochastic case, even starting
with a feasible initial point, the stochastic MM algorithm can no
longer ensure that all the subsequent iterates are still feasible
due to the randomness caused by $\boldsymbol{\xi}$. As a result,
it is much more challenging to design an algorithm for Problem (\ref{eq:mainP})
which involves stochastic non-convex constraints. Indeed, to the best
of our knowledge, there lacks an efficient algorithm in the literature
to handle stochastic non-convex constraints. Most existing algorithms
for non-convex stochastic optimization only consider deterministic
and convex constraints.
\begin{center}
\fbox{\begin{minipage}[t]{0.96\columnwidth}%
\begin{challenge}[Challenges of Algorithm Design]\label{chl:Deterministic-Restriction}Design
an efficient algorithm to find a stationary point of Problem (\ref{eq:mainP})
with stochastic non-convex objective and constraint functions. The
distribution of the random state $\boldsymbol{\xi}$ is not known
a priori and must be obtained from the measurements. Moreover, due
to noisy estimate of the constraints, the sequence of iterates generated
by the algorithm is not always feasible. How to ensure the limiting
point of the algorithm is feasible almost surely? Finally, both the
constraint and objective functions contain expectation and are not
necessarily convex; how to ensure a limiting point of the algorithm
is a stationary point almost surely?\end{challenge}\vspace{-5bp}
\end{minipage}}
\par\end{center}

\subsection{Summary of Algorithm}

We propose a constrained stochastic successive convex approximation
(CSSCA) algorithm to solve Problem (\ref{eq:mainP}), where at each
iteration, $\boldsymbol{x}$ is updated by solving a convex optimization
problem obtained by replacing the objective and constraint functions
$f_{i}\left(\boldsymbol{x}\right),i=0,...,m$ with their convex surrogate
functions $\bar{f}_{i}^{t}\left(\boldsymbol{x}\right),i=0,...,m$. 

Specifically, at iteration $t$, a new realization of the random vector
$\boldsymbol{\xi}^{t}$ is obtained and the surrogate functions $\bar{f}_{i}^{t}\left(\boldsymbol{x}\right),\forall i$
are updated based on $\boldsymbol{\xi}^{t},\boldsymbol{x}^{t}$. The
surrogate function $\bar{f}_{i}^{t}\left(\boldsymbol{x}\right)$ can
be viewed as a convex approximation of $f_{i}\left(\boldsymbol{x}\right)$.
Note that in order to allow maximum freedom for surrogate function
design in different applications, we do not specify the exact form
of the surrogate functions $\bar{f}_{i}^{t}\left(\boldsymbol{x}\right),\forall i$
in this framework algorithm. In Section \ref{subsec:Smooth-Surrogate-Function},
we will give conditions for the surrogate functions $\bar{f}_{i}^{t}\left(\boldsymbol{x}\right),\forall i$
under which the convergence of the algorithm is guaranteed, and a
few common methods to construct the surrogate functions that satisfy
the convergence conditions. 

\begin{figure}
\begin{centering}
\includegraphics[width=85mm]{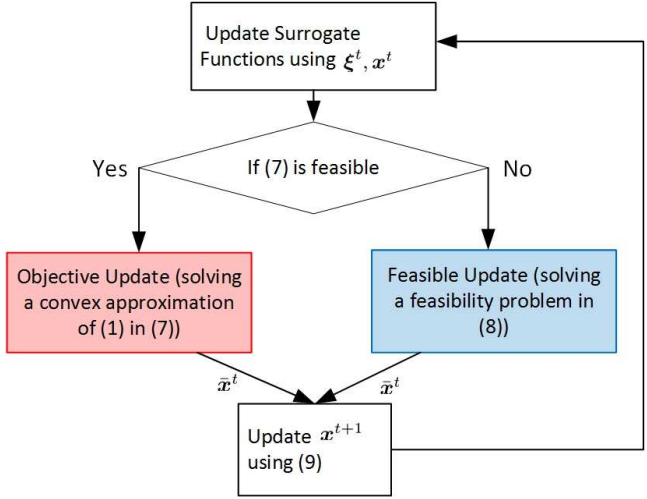}
\par\end{centering}
\caption{\label{fig:Alg1}Block Diagram of CSSCA}
\end{figure}

Then the optimal solution $\bar{\boldsymbol{x}}^{t}$ of the following
problem is solved:
\begin{align}
\bar{\boldsymbol{x}}^{t}=\underset{\boldsymbol{x}\in\mathcal{X}}{\text{argmin}}\: & \bar{f}_{0}^{t}\left(\boldsymbol{x}\right)\label{eq:Pitert}\\
s.t.\: & \bar{f}_{i}^{t}\left(\boldsymbol{x}\right)\leq0,i=1,....,m,\nonumber 
\end{align}
which is a convex approximation of (\ref{eq:mainP}). Note that Problem
(\ref{eq:Pitert}) is not necessarily feasible. If Problem (\ref{eq:Pitert})
turns out to be infeasible, the optimal solution $\bar{\boldsymbol{x}}^{t}$
of the following convex problem is solved: 

\begin{align}
\bar{\boldsymbol{x}}^{t}=\underset{\boldsymbol{x}\in\mathcal{X},\alpha}{\text{argmin}} & \:\alpha\label{eq:Pitert-1}\\
s.t.\: & \bar{f}_{i}^{t}\left(\boldsymbol{x}\right)\leq\alpha,i=1,....,m,\nonumber 
\end{align}
which minimizes the constraint functions. Given $\bar{\boldsymbol{x}}^{t}$
in one of the above two cases, $\boldsymbol{x}$ is updated according
to
\begin{equation}
\boldsymbol{x}^{t+1}=\left(1-\gamma^{t}\right)\boldsymbol{x}^{t}+\gamma^{t}\bar{\boldsymbol{x}}^{t}.\label{eq:updatext}
\end{equation}
where $\left\{ \gamma^{t}\in\left(0,1\right]\right\} $ is a decreasing
sequence satisfying $\gamma^{t}\rightarrow0$, $\sum_{t}\gamma^{t}=\infty$,
$\sum_{t}\left(\gamma^{t}\right)^{2}<\infty$. The overall algorithm
is summarized in Algorithm \ref{alg1} and the block diagram of the
algorithm is given in Fig. \ref{fig:Alg1}.

\begin{algorithm}
\caption{\label{alg1}Constrained stochastic successive convex approximation }

\textbf{Input: }$\left\{ \gamma^{t}\right\} $ satisfying $\gamma^{t}\rightarrow0$,
$\sum_{t}\gamma^{t}=\infty$, $\sum_{t}\left(\gamma^{t}\right)^{2}<\infty$.

\textbf{Initialize:} $\boldsymbol{x}^{0}\in\mathcal{X}$; $t=0$.

\textbf{Step 1: }

The random vector $\boldsymbol{\xi}^{t}$ is realized. 

Update\textbf{ }the surrogate functions $\bar{f}_{i}^{t}\left(\boldsymbol{x}\right),\forall i$
using $\boldsymbol{\xi}^{t},\boldsymbol{x}^{t}$.

\textbf{Step 2: }

//Objective update

\textbf{If} Problem (\ref{eq:Pitert}) is feasible

Solve (\ref{eq:Pitert}) to obtain $\bar{\boldsymbol{x}}^{t}$.

//Feasible update

\textbf{Else }

Solve (\ref{eq:Pitert-1}) to obtain $\bar{\boldsymbol{x}}^{t}$.

\textbf{End if}

\textbf{Step 3: }

Update $\boldsymbol{x}^{t+1}$ according to (\ref{eq:updatext}).

\textbf{Step 4: }

\textbf{Let} $t=t+1$ and return to Step 1.
\end{algorithm}

\subsection{Smooth Surrogate Function Construction\label{subsec:Smooth-Surrogate-Function}}

To guarantee the convergence of Algorithm 1, we need to make the following
assumptions on the surrogate functions. 

\begin{assumption}[Assumptions on properties of surrogate functions]\label{asm:convAnew}For
all $i\in\left\{ 0,...,m\right\} $ and $t=0,1,....$, we have
\begin{enumerate}
\item $\bar{f}_{i}^{t}\left(\boldsymbol{x}\right)$ is uniformly strongly
convex in $\boldsymbol{x}$.
\item $\bar{f}_{i}^{t}\left(\boldsymbol{x}\right)$ is a Lipschitz continuous
function w.r.t. $\boldsymbol{x}$. Moreover, $\limsup_{t_{1},t_{2}\rightarrow\infty}\bar{f}_{i}^{t_{1}}\left(\boldsymbol{x}\right)-\bar{f}_{i}^{t_{2}}\left(\boldsymbol{x}\right)-B\left\Vert \boldsymbol{x}^{t_{1}}-\boldsymbol{x}^{t_{2}}\right\Vert \leq0,\forall\boldsymbol{x}\in\mathcal{X}$
for some constant $B>0$.
\item For any $\boldsymbol{x}\in\mathcal{X}$, the function $\bar{f}_{i}^{t}\left(\boldsymbol{x}\right)$,
its derivative, and its second order derivative are uniformly bounded.
\end{enumerate}
\end{assumption}

\begin{assumption}[Asymptotic consistency of surrogate functions]\label{asm:convBnew}For
all $i\in\left\{ 0,...,m\right\} $, we have 
\begin{align*}
\lim_{t\rightarrow\infty}\left|\bar{f}_{i}^{t}\left(\boldsymbol{x}^{t}\right)-f_{i}\left(\boldsymbol{x}^{t}\right)\right| & =0,\\
\lim_{t\rightarrow\infty}\left\Vert \nabla\bar{f}_{i}^{t}\left(\boldsymbol{x}^{t}\right)-\nabla f_{i}\left(\boldsymbol{x}^{t}\right)\right\Vert  & =0.
\end{align*}

\end{assumption}

These assumptions are quite standard and are satisfied for a large
class of surrogate functions. In the following, we give some common
examples of surrogate functions $\bar{f}_{i}^{t}\left(\boldsymbol{x}\right)$
that satisfy the above assumptions.

\subsubsection{Recursive Surrogate Function}

In this case, the surrogate function $\bar{f}_{i}^{t}\left(\boldsymbol{x}\right)$
can be expressed using a recursive formula as 
\begin{equation}
\bar{f}_{i}^{t}\left(\boldsymbol{x}\right)=\left(1-\rho^{t}\right)\bar{f}_{i}^{t-1}\left(\boldsymbol{x}\right)+\rho^{t}\hat{g}_{i}\left(\boldsymbol{x},\boldsymbol{x}^{t},\boldsymbol{\xi}^{t}\right),\label{eq:TSF}
\end{equation}
where $\rho^{t}\in\left(0,1\right]$ is a sequence to be properly
chosen, $\hat{g}_{i}\left(\boldsymbol{x},\boldsymbol{x}^{t},\boldsymbol{\xi}^{t}\right)$
is a convex approximation of the function $g_{i}\left(\boldsymbol{x},\boldsymbol{\xi}^{t}\right)$
around the point $\boldsymbol{x}^{t}$ and it is called the \textit{sample
surrogate function} at the $t$-th iteration. The initial value $\bar{f}_{i}^{-1}\left(\boldsymbol{x}\right)=0$. 

\begin{assumption}[Assumptions on $\hat{g}_{i}\left(\boldsymbol{x},\boldsymbol{x}^{t},\boldsymbol{\xi}^{t}\right)$]\label{asm:convRSF}For
all $i\in\left\{ 0,...,m\right\} $, we have
\begin{enumerate}
\item $\hat{g}_{i}\left(\boldsymbol{x},\boldsymbol{x},\boldsymbol{\xi}\right)=g_{i}\left(\boldsymbol{x},\boldsymbol{\xi}\right)$
and $\nabla\hat{g}_{i}\left(\boldsymbol{x},\boldsymbol{x},\boldsymbol{\xi}\right)=\nabla g_{i}\left(\boldsymbol{x},\boldsymbol{\xi}\right),\:\forall\boldsymbol{x}\in\mathcal{X},\forall\boldsymbol{\xi}\in\Omega$.
\item $\hat{g}_{i}\left(\boldsymbol{x},\boldsymbol{y},\boldsymbol{\xi}\right)$
is strongly convex in $\boldsymbol{x}$ for all $\boldsymbol{y}\in\mathcal{X},\boldsymbol{\xi}\in\Omega$.
\item For any $\boldsymbol{\xi}\in\Omega$ and $\boldsymbol{y}\in\mathcal{X}$,
the function $\hat{g}_{i}\left(\boldsymbol{x},\boldsymbol{y},\boldsymbol{\xi}\right)$
is Lipschitz continuous in both $\boldsymbol{x}$ and $\boldsymbol{y}$.
\item The function $\hat{g}_{i}\left(\boldsymbol{x},\boldsymbol{y},\boldsymbol{\xi}\right)$,
its derivative, and its second order derivative w.r.t. $\boldsymbol{x}$
are uniformly bounded.
\end{enumerate}
\end{assumption}

An example of first order sample surrogate function $\hat{g}_{i}\left(\boldsymbol{x},\boldsymbol{y},\boldsymbol{\xi}\right)$
that satisfies Assumption \ref{asm:convRSF} is
\begin{equation}
\hat{g}_{i}\left(\boldsymbol{x},\boldsymbol{y},\boldsymbol{\xi}\right)=g_{i}\left(\boldsymbol{y},\boldsymbol{\xi}\right)+\nabla^{T}g_{i}\left(\boldsymbol{y},\boldsymbol{\xi}\right)\left(\boldsymbol{x}-\boldsymbol{y}\right)+\tau_{i}\left\Vert \boldsymbol{x}-\boldsymbol{y}\right\Vert ^{2},\label{eq:ghead}
\end{equation}
where $\tau_{i}>0$ can be any constant, and the term $\tau_{i}\left\Vert \boldsymbol{x}-\boldsymbol{y}\right\Vert ^{2}$
is used to ensure strong convexity. The surrogate function in (\ref{eq:ghead})
includes the Lipschitz gradient surrogate function in \cite{NIPS2013_5129_StochasticMM}
for stochastic MM as a special case. In the Lipschitz gradient surrogate
function, $\tau_{i}$ must be sufficiently large to ensure that $\hat{g}_{i}\left(\boldsymbol{x},\boldsymbol{y},\boldsymbol{\xi}\right)\geq g_{i}\left(\boldsymbol{x},\boldsymbol{\xi}\right),\:\forall\boldsymbol{x}\in\mathcal{X}$.
However, (\ref{eq:ghead}) does not have such a restriction and thus
provides more freedom to design better surrogate functions. 

\subsubsection{Structured Surrogate Function in \cite{Yang_TSP2016_SSCA}}

Suppose $g_{i}\left(\boldsymbol{x},\xi\right)$ can be divided into
two components as
\[
g_{i}\left(\boldsymbol{x},\xi\right)=g_{i}^{c}\left(\boldsymbol{x},\xi\right)+g_{i}^{\bar{c}}\left(\boldsymbol{x},\xi\right),
\]
where $g_{i}^{c}\left(\boldsymbol{x},\xi\right)$ is convex and $g_{i}^{\bar{c}}\left(\boldsymbol{x},\xi\right)$
can be either convex or non-convex. Then the structured surrogate
function $\bar{f}_{i}^{t}\left(\boldsymbol{x}\right)$ is given by
\cite{Yang_TSP2016_SSCA}
\begin{align}
\bar{f}_{i}^{t}\left(\boldsymbol{x}\right) & =\left(1-\rho^{t}\right)f_{i}^{t-1}+\rho^{t}g_{i}^{c}\left(\boldsymbol{x},\xi^{t}\right)\nonumber \\
 & +\rho^{t}g_{i}^{\bar{c}}\left(\boldsymbol{x}^{t},\xi^{t}\right)+\rho^{t}\nabla^{T}g_{i}^{\bar{c}}\left(\boldsymbol{x}^{t},\xi^{t}\right)\left(\boldsymbol{x}-\boldsymbol{x}^{t}\right)\nonumber \\
 & +\left(1-\rho^{t}\right)\left(\mathbf{f}_{i}^{t-1}\right)^{T}\left(\boldsymbol{x}-\boldsymbol{x}^{t}\right)+\tau_{i}\left\Vert \boldsymbol{x}-\boldsymbol{x}^{t}\right\Vert ^{2},\label{eq:SSF}
\end{align}
where $\tau_{i}>0$ can be any constant, $f_{i}^{t}$ is an approximation
for $\mathbb{E}\left[g_{i}\left(\boldsymbol{x}^{t},\xi\right)\right]$
and it is updated recursively according to
\[
f_{i}^{t}=\left(1-\rho^{t}\right)f_{i}^{t-1}+\rho^{t}g_{i}\left(\boldsymbol{x}^{t},\xi^{t}\right),
\]
with $f_{i}^{-1}=0$, and $\mathbf{f}_{i}^{t}$ is an approximation
for the gradient $\nabla\mathbb{E}\left[g_{i}\left(\boldsymbol{x}^{t},\xi\right)\right]$,
which is updated recursively according to
\[
\mathbf{f}_{i}^{t}=\left(1-\rho^{t}\right)\mathbf{f}_{i}^{t-1}+\rho^{t}\nabla g_{i}\left(\boldsymbol{x}^{t},\xi^{t}\right),
\]
with $\mathbf{f}_{i}^{-1}=\boldsymbol{0}$. The structured surrogate
function in (\ref{eq:SSF}) contains the convex component $g_{i}^{c}\left(\boldsymbol{x},\xi\right)$
of the original sample objective function $g_{i}\left(\boldsymbol{x},\xi\right)$,
which helps to reduce the approximation error and potentially achieve
a faster initial convergence speed \cite{Yang_TSP2016_SSCA}.

\subsubsection{Validity of the above Surrogate Functions}

We formally prove that the above two surrogate functions satisfy the
conditions in Assumptions \ref{asm:convAnew} and \ref{asm:convBnew},
under the following conditions on the step sizes.

\begin{assumption}[Assumptions on step sizes]\label{asm:convS}$\:$
\begin{enumerate}
\item $\rho^{t}\rightarrow0$, $\sum_{t}\rho^{t}=\infty$, $\sum_{t}\left(\rho^{t}\right)^{2}<\infty$,
\item $\lim_{t\rightarrow\infty}\gamma^{t}/\rho^{t}=0$.
\end{enumerate}
\end{assumption}

A typical choice of $\rho^{t},\gamma^{t}$ that satisfies Assumption
\ref{asm:convS} is $\rho^{t}=O\left(t^{-\kappa_{1}}\right)$, $\gamma^{t}=O\left(t^{-\kappa_{2}}\right)$,
where $0.5<\kappa_{1}<\kappa_{2}\leq1$. Such form of step sizes have
been widely considered in stochastic optimization \cite{Yang_TSP2016_SSCA}.
\begin{prop}
[Validity of the recursive surrogate]\label{prop:Validity-of-theRSF}Under
Assumption \ref{asm:convP}, \ref{asm:convRSF} and \ref{asm:convS},
if we choose the surrogate functions $\bar{f}_{i}^{t}\left(\boldsymbol{x}\right),\forall i$
as in (\ref{eq:TSF}), then Assumption \ref{asm:convAnew} and \ref{asm:convBnew}
are satisfied.
\end{prop}

Please refer to Appendix \ref{subsec:Proof-of-Proposition} for the
proof.
\begin{prop}
[Validity of the structured surrogate]Under Assumption \ref{asm:convS},
if we choose the surrogate functions $\bar{f}_{i}^{t}\left(\boldsymbol{x}\right),\forall i$
as in (\ref{eq:SSF}), then Assumptions \ref{asm:convAnew} and \ref{asm:convBnew}
are satisfied.
\end{prop}

The proof is similar to that of Proposition \ref{prop:Validity-of-theRSF}
and is omitted for conciseness.

Note that Assumptions 1 - 3 are the key assumptions used to establish
the convergence of the algorithm, while Assumption 4 and 5 are only
used to ensure that the above two example surrogate functions satisfy
the general condition in Assumption 3.

\subsection{Key Differences from the Conventional Stochastic SCA}

The conventional stochastic SCA algorithms in \cite{NIPS2013_5129_StochasticMM,Yang_TSP2016_SSCA}
only consider deterministic and convex constraints. There are two
key differences between the conventional stochastic SCA and the proposed
CSSCA due to the consideration of stochastic non-convex constraints. 

First, in the conventional stochastic SCA, the constraints are deterministic
and convex. As a result, there is no need to construct and update
the surrogate functions for constraints. In CSSCA, however, we need
to construct and update the surrogate functions for constraints. 

Second, the sequence of iterates generated by the conventional stochastic
SCA is always feasible. In contrast, the sequence of iterates generated
by the CSSCA may not be feasible, and thus it is necessary to perform
the feasible update by solving (\ref{eq:Pitert-1}) to ensure that
the algorithm converges to a feasible point. Specifically, in Step
2 of CSSCA, when Problem (\ref{eq:Pitert}) is feasible, we do an
objective update by solving a convex approximation of (\ref{eq:mainP})
in (\ref{eq:Pitert}), aiming at reducing the objective function.
Otherwise, we do a feasible update by solving an approximate feasibility
problem in (\ref{eq:Pitert-1}), aiming at reducing the constraint
functions.

In summary, due to the stochastic non-convex constraints, the sequence
of iterates generated by the CSSCA may not be feasible and we have
to do a feasible update as well. As a result, the convergence analysis
of the CSSCA is also more challenging than that of the conventional
stochastic SCA. We shall provide the convergence proof in the next
section.
\begin{rem}
The proposed CSSCA algorithm can be easily tailored to solve a deterministic
non-convex constrained problem (i.e., $f_{i}\left(\boldsymbol{x}\right)\triangleq g_{i}\left(\boldsymbol{x},\xi\right),\forall i$
for a deterministic system state $\xi$), by choosing the surrogate
function to be the sample surrogate function, i.e., $\bar{f}_{i}^{t}\left(\boldsymbol{x}\right)=\hat{g}_{i}\left(\boldsymbol{x},\boldsymbol{x}^{t},\boldsymbol{\xi}\right),\forall i$.
In this case, we have $\bar{f}_{i}^{t}\left(\boldsymbol{x}^{t}\right)=f_{i}\left(\boldsymbol{x}^{t}\right),\forall t$
and $\nabla\bar{f}_{i}^{t}\left(\boldsymbol{x}^{t}\right)=\nabla f_{i}\left(\boldsymbol{x}^{t}\right),\forall t$,
i.e., the convergence of surrogate function is achieved at each iteration
since there is no randomness caused by the random system state $\xi$.
Therefore, the convergence speed of the deterministic version of the
CSSCA algorithm is usually faster than that of the stochastic version. 
\end{rem}

\section{Convergence Analysis\label{sec:Convergence-Analysis}}

There are several challenges in the convergence proof for Algorithm
1, as explained below. 

\fbox{\begin{minipage}[t]{0.96\columnwidth}%
\begin{challenge}[Challenges of Convergence Proof]\label{chl:Deterministic-Restriction-1}We
need to show that at every limiting point, all constraints are satisfied,
which is non-trivial since Algorithm 1 may oscillate between the feasible
update and objective update. Moreover, the limiting point is obtained
by averaging over all the previous outputs from either feasible updates
or objective updates, which makes it difficult to show that the limiting
point is a stationary point of the original problem (\ref{eq:mainP}).\end{challenge}%
\end{minipage}}\vspace{2bp}

To state the convergence result, we need to prove the convergence
of surrogate functions, and introduce the concept of Slater condition
for the converged surrogate functions.
\begin{lem}
[Convergence of the surrogate functions]\label{lem:Convergence-surrogate}Suppose
Assumptions \ref{asm:convP}, \ref{asm:convAnew} and \ref{asm:convBnew}
are satisfied. Consider a subsequence $\left\{ \boldsymbol{x}^{t_{j}}\right\} _{j=1}^{\infty}$
converging to a limit point $\boldsymbol{x}^{*}$. There exist uniformly
continuous functions $\hat{f}_{i}\left(\boldsymbol{x}\right)$ such
that
\begin{align}
\lim_{j\rightarrow\infty}\bar{f}_{i}^{t_{j}}\left(\boldsymbol{x}\right) & =\hat{f}_{i}\left(\boldsymbol{x}\right),\:\forall\boldsymbol{x}\in\mathcal{X},\label{eq:ghfhead}
\end{align}
almost surely. Moreover, we have 
\begin{align}
\left|\hat{f}_{i}\left(\boldsymbol{x}^{*}\right)-f_{i}\left(\boldsymbol{x}^{*}\right)\right| & =0,\nonumber \\
\left\Vert \nabla\hat{f}_{i}\left(\boldsymbol{x}^{*}\right)-\nabla f_{i}\left(\boldsymbol{x}^{*}\right)\right\Vert  & =0.\label{eq:convsur}
\end{align}
\end{lem}

Please refer to Appendix \ref{subsec:Proof-of-Lemmaconvf} for the
proof.

\textbf{Slater condition for the converged surrogate functions:} Given
a subsequence $\left\{ \boldsymbol{x}^{t_{j}}\right\} _{j=1}^{\infty}$
converging to a limit point $\boldsymbol{x}^{*}$ and let $\hat{f}_{i}\left(\boldsymbol{x}\right),\forall i$
be the converged surrogate functions as defined in Lemma \ref{lem:Convergence-surrogate}.
We say that the Slater condition is satisfied at $\boldsymbol{x}^{*}$
if there exists $\boldsymbol{x}\in\textrm{relint}\mathcal{X}$ such
that
\[
\hat{f}_{i}\left(\boldsymbol{x}\right)<0,\:\forall i=1,...,m.
\]
A similar Slater condition is also assumed in \cite{Meisam_thesis14_BSUM}
to prove the convergence of a deterministic MM algorithm with non-convex
constraints.

With the Lemma \ref{lem:Convergence-surrogate} and Slater condition,
we are ready to prove the following main convergence result.
\begin{thm}
[Convergence of Algorithm 1]\label{thm:Convergence-of-Algorithm1-1}Suppose
Assumptions \ref{asm:convP}, \ref{asm:convAnew} and \ref{asm:convBnew}
are satisfied, and the initial point $\boldsymbol{x}^{0}$ is a feasible
point, i.e., $\max_{i\in\left\{ 1,...,m\right\} }f_{i}\left(\boldsymbol{x}^{0}\right)\leq0$.
Let $\left\{ \boldsymbol{x}^{t}\right\} _{t=1}^{\infty}$ denote the
iterates generated by Algorithm 1 with a sufficiently small initial
step size $\gamma^{0}$. Then every limiting point $\boldsymbol{x}^{*}$
of $\left\{ \boldsymbol{x}^{t}\right\} _{t=1}^{\infty}$ satisfying
the Slater condition is a stationary point of Problem (\ref{eq:mainP})
almost surely.
\end{thm}

Please refer to Appendix \ref{subsec:Proof-of-Theorem-local} for
the proof. The above convergence result states that, starting from
a feasible initial point, Algorithm 1 will converge to a stationary
point almost surely, providing that the step sizes are sufficiently
small (since we assume $\gamma^{t}$ is a decreasing sequence, a sufficiently
small initial step size\textit{ $\gamma^{0}$} implies all step sizes
are sufficiently small), and the Slater condition is satisfied. Note
that due to the stochastic nature of the problem/algorithm, we need
to assume that the step size $\gamma^{t}$ is sufficiently small to
make it easier to handle the randomness caused by the random system
state for tractable convergence analysis and rigorous convergence
proof. However, choosing a small $\gamma^{0}$ is usually not mandatory
for the practical convergence of Algorithm 1. In the simulations,
we find that the algorithm can still converge even when the initial
step size $\gamma^{0}$ is not small. In fact, in practice, we may
prefer to choose a not very small $\gamma^{0}$ to achieve a faster
initial convergence speed.

Finally, we discuss the convergence behavior of Algorithm 1 with an
infeasible initial point. In this case, it follows from the analysis
in Appendix \ref{subsec:Proof-of-Theorem-local} that Algorithm 1
either converges to stationary points of Problem (\ref{eq:mainP}),
or converges to the following \textit{undesired set}:
\[
\overline{\mathcal{X}}_{C}^{*}=\left\{ \boldsymbol{x}:\:f\left(\boldsymbol{x}\right)>0,\:\boldsymbol{x}\in\mathcal{X}_{C}^{*}\right\} ,
\]
where $\mathcal{X}_{C}^{*}$ is the set of stationary points of the
following constraint minimization problem:
\begin{align}
\mathcal{P}_{C}:\:\min_{\boldsymbol{x}\in\mathcal{X}}\: & f\left(\boldsymbol{x}\right)\triangleq\max_{i\in\left\{ 1,...,m\right\} }f_{i}\left(\boldsymbol{x}\right).\label{eq:FP-2}
\end{align}
Due to the proposed feasible update, Algorithm 1 may still converge
to a stationary point of Problem (\ref{eq:mainP}) even when the initial
point is infeasible, as long as the initial point is not close to
an undesired point $\boldsymbol{x}_{C}^{*}\in\overline{\mathcal{X}}_{C}^{*}$
such that the algorithm gets stuck in this undesired point. In practice,
if we run Algorithm 1 with multiple random initial points, it is likely
that the algorithm with one of the initial points will converge to
a stationary point of Problem (\ref{eq:mainP}).
\begin{rem}
In CSSCA, we can also use multiple samples of system state to calculate
the surrogate functions at each iteration. As long as Assumption 3
is satisfied, the convergence of CSSCA is still guaranteed. Using
multiple system state samples at each iteration can reduce the randomness
of surrogate functions and thus potentially reduce the number of iterations
required to converge, but the complexity per iteration will also increase.
Therefore, the proposed CSSCA has the freedom to control the tradeoff
between the number of iterations and the complexity per iteration.
\end{rem}

\section{Parallel Implementation for Decoupled Constraints\label{sec:Parallel-Implementation-for}}

In this section, we consider a parallel implementation of Algorithm
1 over a distributed system for stochastic optimization problems with
decoupled constraints. There are $K$ nodes in the system. The optimization
variables are partitioned into $K$ blocks $\boldsymbol{x}=\left(\boldsymbol{x}_{k}\right)_{k=1}^{K}$
and node $k$ needs to optimize the $k$-th block $\boldsymbol{x}_{k}$.
Specifically, the stochastic optimization problem with decoupled constraints
is formulated as
\begin{align}
\min_{\boldsymbol{x}\triangleq\left(\boldsymbol{x}_{k}\right)_{k=1}^{K}}\: & f_{0}\left(\boldsymbol{x}\right)\triangleq\mathbb{E}\left[g_{0}\left(\boldsymbol{x},\xi\right)\right]\label{eq:maindistP}\\
s.t.\: & f_{i,k}\left(\boldsymbol{x}_{k}\right)\triangleq\mathbb{E}\left[g_{i,k}\left(\boldsymbol{x}_{k},\xi\right)\right]\leq0,\nonumber \\
 & i=1,....,m_{k},k=1,...,K.\nonumber 
\end{align}
In Problem (\ref{eq:maindistP}) , there are $K$ groups of constraints,
where the $k$-th constraint group contains $m_{k}$ constraints with
the constraint functions $f_{i,k}\left(\boldsymbol{x}_{k}\right),i=1,...,m_{k}$
only depending on the $k$-th block $\boldsymbol{x}_{k}$. Problem
(\ref{eq:maindistP}) includes many distributed optimization problems,
such as the multi-agent optimization problems considered in \cite{Yang_TSP2016_SSCA},
as special cases.

We use the recursive surrogate function in (\ref{eq:TSF}) as an example
to illustrate the parallel implementation of Algorithm 1. The parallel
implementation for the structured surrogate function is similar. In
this case, the sample surrogate function for each function $g_{i,k}\left(\boldsymbol{x}_{k},\xi\right)$
in the constraint in (\ref{eq:maindistP}) is denoted by $\hat{g}_{i,k}\left(\boldsymbol{x}_{k},\boldsymbol{x}_{k}^{t},\boldsymbol{\xi}^{t}\right)$,
which is naturally decoupled over the $K$ blocks $\left(\boldsymbol{x}_{k}\right)_{k=1}^{K}$.
To facilitate parallel implementation of Algorithm 1, we consider
the\textit{ decoupled sample surrogate function} for the function
$g_{0}\left(\boldsymbol{x},\xi\right)$ in the objective, which has
the following form:
\[
\hat{g}_{0}\left(\boldsymbol{x},\boldsymbol{x}^{t},\boldsymbol{\xi}^{t}\right)=\sum_{k=1}^{K}\hat{g}_{0,k}\left(\boldsymbol{x}_{k},\boldsymbol{x}^{t},\boldsymbol{\xi}^{t}\right).
\]
One example of the decoupled sample surrogate function is 
\begin{align*}
\hat{g}_{0,k}\left(\boldsymbol{x}_{k},\boldsymbol{x}^{t},\boldsymbol{\xi}^{t}\right) & =\frac{1}{K}g_{0}\left(\boldsymbol{x}^{t},\boldsymbol{\xi}^{t}\right)+\nabla_{\boldsymbol{x}_{k}}^{T}g_{0}\left(\boldsymbol{x}^{t},\boldsymbol{\xi}^{t}\right)\left(\boldsymbol{x}_{k}-\boldsymbol{x}_{k}^{t}\right)\\
 & +\tau_{k}\left\Vert \boldsymbol{x}_{k}-\boldsymbol{x}_{k}^{t}\right\Vert ^{2},\forall k
\end{align*}
where $\tau_{k}>0$ is some constant. 

By choosing a decoupled sample surrogate function for $g_{0}\left(\boldsymbol{x},\xi\right)$,
the surrogate function $\bar{f}_{0}^{t}\left(\boldsymbol{x}\right)$
for the objective $f_{0}\left(\boldsymbol{x}\right)$ is given by
\[
\bar{f}_{0}^{t}\left(\boldsymbol{x}\right)=\sum_{k=1}^{K}\bar{f}_{0,k}^{t}\left(\boldsymbol{x}_{k}\right),
\]
where 
\[
\bar{f}_{0,k}^{t}\left(\boldsymbol{x}_{k}\right)=\left(1-\rho^{t}\right)\bar{f}_{0,k}^{t-1}\left(\boldsymbol{x}_{k}\right)+\rho^{t}\hat{g}_{0,k}\left(\boldsymbol{x}_{k},\boldsymbol{x}^{t},\boldsymbol{\xi}^{t}\right),
\]
with $\bar{f}_{0,k}^{-1}\left(\boldsymbol{x}_{k}\right)=0$. The surrogate
function $\bar{f}_{i,k}^{t}\left(\boldsymbol{x}_{k}\right)$ for the
$i$-th constraint in the $k$-th constraint group is given by
\begin{align*}
\bar{f}_{i,k}^{t}\left(\boldsymbol{x}_{k}\right)= & \left(1-\rho^{t}\right)\bar{f}_{i,k}^{t-1}\left(\boldsymbol{x}_{k}\right)+\rho^{t}\hat{g}_{i,k}\left(\boldsymbol{x}_{k},\boldsymbol{x}^{t},\boldsymbol{\xi}^{t}\right),
\end{align*}
with $\bar{f}_{i,k}^{-1}\left(\boldsymbol{x}_{k}\right)=0$. Note
that in the surrogate update step (Step 1 of Algorithm 1), the surrogate
functions $\bar{f}_{i,k}^{t}\left(\boldsymbol{x}_{k}\right),i=0,1,...,m_{k}$
corresponding to the $k$-th block $\boldsymbol{x}_{k}$ can be performed
distributedly at node $k$.

In the objective update in Step 2, the optimization problem in (\ref{eq:Pitert})
can be decoupled into $K$ independent subproblems as
\begin{align}
\bar{\boldsymbol{x}}_{k}^{t}=\underset{\boldsymbol{x}_{k}}{\text{argmin}\:} & \bar{f}_{0,k}^{t}\left(\boldsymbol{x}_{k}\right)\label{eq:Pitert-dist}\\
s.t.\:\bar{f}_{i,k}^{t}\left(\boldsymbol{x}_{k}\right) & \leq0,i=1,....,m_{k},\nonumber 
\end{align}
for $k=1,...,K$, which can be solved by the $K$ nodes in a distributed
and parallel way. Similarly, in the constraint update in Step 2, the
optimization problem in (\ref{eq:Pitert-1}) can be decoupled into
$K$ independent subproblems as
\begin{align}
\bar{\boldsymbol{x}}_{k}^{t}=\underset{\boldsymbol{x}_{k},\alpha_{k}}{\text{argmin}}\: & \alpha_{k}\label{eq:Pitert-dist-1}\\
s.t.\:\bar{f}_{i,k}^{t}\left(\boldsymbol{x}_{k}\right) & \leq\alpha_{k},i=1,....,m_{k},\nonumber 
\end{align}
for $k=1,...,K$, which can be solved by the $K$ nodes in a distributed
and parallel way. The optimal solution of (\ref{eq:Pitert-1}) is
given by $\bar{\boldsymbol{x}}^{t}=\left(\bar{\boldsymbol{x}}_{k}^{t}\right)_{k=1}^{K}$
and the optimal value of (\ref{eq:Pitert-1}) is given by $\alpha=\min_{k}\alpha_{k}$.
The update of $\boldsymbol{x}$ in Step 3 is also decoupled as
\begin{equation}
\boldsymbol{x}_{k}^{t+1}=\left(1-\gamma^{t}\right)\boldsymbol{x}_{k}^{t}+\gamma^{t}\bar{\boldsymbol{x}}_{k}^{t}.\label{eq:updatext-1}
\end{equation}

\section{Applications\label{sec:Applications}}

In this section, we shall apply the proposed CSSCA to solve the three
application problems described in Section \ref{sec:System-Model}.
As discussed in the introduction, there are only a few algorithms
that can handle the non-convex stochastic constraints. Among them,
sample average approximation (SAA) is a common method to solve a general
stochastic optimization problem with non-convex stochastic constraints
\cite{SPlecbook}. However, the SAA method needs to collect a large
number of samples for the random state before solving the stochastic
optimization problem. Therefore, it requires more memory to store
the samples and the computational complexity is also higher than the
proposed CSSCA. Moreover, the computational complexity is also lower.
The online primal-dual algorithm in \cite{Mahdavi2013Online} may
also be used to solve a non-convex stochastic optimization problem,
although the convergence is not guaranteed. On the other hand, the
Bernstein approximation and its variations \cite{Wang_ICASSP11_Bernstein-type}
are the state-of-the-art algorithms to handle the chance constraint
in Example \ref{exa:Portfolio-optimization}. Therefore, we compare
the performance of the CSSCA with the SAA and online primal-dual (for
Example \ref{exa:Robust-MIMO-Transmit} and \ref{exa:Massive-MIMO-Hybrid}),
as well as the Bernstein approximation (for Example \ref{exa:Portfolio-optimization}).
The stepsizes/parameters in all algorithms are tuned such that they
can achieve their best empirical convergence speed. The simulation
results clearly show the advantage of the proposed CSSCA over these
baseline algorithms.

\subsection{MIMO Transmit Signal Design with Imperfect CSI}

Consider the MIMO transmit signal design problem with imperfect CSI
as in (\ref{eq:RBF}). The objective function is a linear deterministic
convex function, and the constraints can be rewritten as $\mathbb{E}\left[g_{k}\left(\boldsymbol{Q},\boldsymbol{H}\right)\right]\leq0,\forall k$
with
\begin{align*}
g_{k}\left(\boldsymbol{Q},\boldsymbol{H}\right) & =g_{k}^{c}\left(\boldsymbol{Q},\boldsymbol{H}\right)+g_{k}^{\bar{c}}\left(\boldsymbol{Q},\boldsymbol{H}\right),\\
g_{k}^{c}\left(\boldsymbol{Q},\boldsymbol{H}\right) & =r_{k}-\log\left(\sum_{j=1}^{K}\boldsymbol{h}_{k}^{H}\boldsymbol{Q}_{j}\boldsymbol{h}_{k}+\sigma_{k}^{2}\right),\\
g_{k}^{\bar{c}}\left(\boldsymbol{Q},\boldsymbol{H}\right) & =\log\left(\sum_{j\neq k}\boldsymbol{h}_{k}^{H}\boldsymbol{Q}_{j}\boldsymbol{h}_{k}+\sigma_{k}^{2}\right),
\end{align*}
where $\boldsymbol{Q}=\left\{ \boldsymbol{Q}_{i}\right\} _{i=1}^{K}$
is the set of all covariance matrices, and $\boldsymbol{H}=\left[\boldsymbol{h}_{k}\right]_{k=1,...,K}^{H}\in\mathbb{C}^{K\times n}$
is the composite channel matrix. Note that $g_{k}^{c}\left(\boldsymbol{Q},\boldsymbol{H}\right)$
and $g_{k}^{\bar{c}}\left(\boldsymbol{Q},\boldsymbol{H}\right)$ are
the convex and non-convex components, respectively, of $g_{k}\left(\boldsymbol{Q},\boldsymbol{H}\right)$.
This motivates us to choose a structured surrogate function. Specifically,
we first calculate the gradient of the non-convex component with respect
to $\boldsymbol{Q}_{i}$ as
\[
\nabla_{\boldsymbol{Q}_{i}}g_{k}^{\bar{c}}\left(\boldsymbol{Q},\boldsymbol{H}\right)=\frac{\boldsymbol{h}_{k}\boldsymbol{h}_{k}^{H}}{\sum_{j\neq k}\boldsymbol{h}_{k}^{H}\boldsymbol{Q}_{j}\boldsymbol{h}_{k}+\sigma_{k}^{2}},\forall i\neq k,
\]
and $\nabla_{\boldsymbol{Q}_{k}}g_{k}^{\bar{c}}\left(\boldsymbol{Q},\boldsymbol{H}\right)=\boldsymbol{0}$,
and the gradient of the convex component with respect to $\boldsymbol{Q}_{i}$
as

\[
\nabla_{\boldsymbol{Q}_{i}}g_{k}^{c}\left(\boldsymbol{Q},\boldsymbol{H}^{t}\right)=-\frac{\boldsymbol{h}_{k}\boldsymbol{h}_{k}^{H}}{\sum_{j=1}^{K}\boldsymbol{h}_{k}^{H}\boldsymbol{Q}_{j}\boldsymbol{h}_{k}+\sigma_{k}^{2}},\forall i.
\]
Then the surrogate function is given by
\begin{align}
\bar{f}_{k}^{t}\left(\boldsymbol{Q}\right) & =\left(1-\rho^{t}\right)f_{k}^{t-1}+\rho^{t}g_{k}^{c}\left(\boldsymbol{Q},\boldsymbol{H}^{t}\right)+\rho^{t}g_{k}^{\bar{c}}\left(\boldsymbol{Q}^{t},\boldsymbol{H}^{t}\right)\nonumber \\
 & +\rho^{t}\sum_{i\neq k}\mathfrak{R}\left[Tr\left(\nabla_{\boldsymbol{Q}_{i}}^{H}g_{k}^{\bar{c}}\left(\boldsymbol{Q}^{t},\boldsymbol{H}^{t}\right)\left(\boldsymbol{Q}_{i}-\boldsymbol{Q}_{i}^{t}\right)\right)\right]\nonumber \\
 & +\left(1-\rho^{t}\right)\sum_{i=1}^{K}\mathfrak{R}\left[Tr\left(\left(\mathbf{F}_{i}^{t-1}\right)^{H}\left(\boldsymbol{Q}_{i}-\boldsymbol{Q}_{i}^{t}\right)\right)\right]\nonumber \\
 & +\tau_{k}\sum_{i=1}^{K}Tr\left(\left(\boldsymbol{Q}_{i}-\boldsymbol{Q}_{i}^{t}\right)\left(\boldsymbol{Q}_{i}-\boldsymbol{Q}_{i}^{t}\right)^{H}\right),\label{eq:sfexm1}
\end{align}
where $\mathfrak{R}\left[\cdot\right]$ is the real operator, $Tr\left(\cdot\right)$
is the trace operator, $\boldsymbol{H}^{t}=\left[\boldsymbol{h}_{k}^{t}\right]_{k=1,...,K}^{H}\in\mathbb{C}^{K\times n}$
with $\boldsymbol{h}_{k}^{t}=\hat{\boldsymbol{h}}_{k}+\boldsymbol{e}_{k}^{t}$,
and $\boldsymbol{e}_{k}^{t},k=1,...,K$ denotes the channel estimation
error observed (generated) at iteration $t$. The matrices $\mathbf{F}_{i}^{t-1}$
can be calculated recursively as 
\[
\mathbf{F}_{i}^{t}=\left(1-\rho^{t}\right)\mathbf{F}_{i}^{t-1}+\rho^{t}\nabla_{\boldsymbol{Q}_{i}}g_{k}\left(\boldsymbol{Q}^{t},\boldsymbol{H}^{t}\right),
\]
where $\nabla_{\boldsymbol{Q}_{i}}g_{k}\left(\boldsymbol{Q}^{t},\boldsymbol{H}^{t}\right)=\nabla_{\boldsymbol{Q}_{i}}g_{k}^{\bar{c}}\left(\boldsymbol{Q}^{t},\boldsymbol{H}^{t}\right)+\nabla_{\boldsymbol{Q}_{i}}g_{k}^{c}\left(\boldsymbol{Q}^{t},\boldsymbol{H}^{t}\right)$,
and the constant $f_{k}^{t}$ can be calculated as
\[
f_{k}^{t}=\frac{1}{t}\sum_{j=1}^{t}g_{k}\left(\boldsymbol{Q}^{t},\boldsymbol{H}^{j}\right).
\]
With the surrogate functions in (\ref{eq:sfexm1}), we can implement
the proposed CSSCA for Problem (\ref{eq:RBF}).

We compare the proposed CSSCA with the SAA and online primal-dual
algorithms. After applying the SAA on the constraint functions using
$N=200$ realizations of channel estimation errors, the problem becomes
a deterministic optimization problem with non-convex constraints.
We apply the deterministic SCA method in \cite{Meisam_thesis14_BSUM}
to solve the resulting non-convex problem. Similarly, the SAA of the
constraint function also consists of a convex component plus a concave
component, and in the deterministic SCA, only approximation for the
concave component is required. Specifically, we use linear approximation
(i.e., first order Taylor expansion) as the surrogate function for
the concave component in the deterministic SCA method. In both CSSCA
and ``SAA + SCA'', CVX \cite{cvx} is used to solve the convex subproblem
at each iteration.

\textit{Numerical Results:} In the simulations, there are $n=8$ antennas
and $K=4$ users. The estimated channel coefficients $\hat{\boldsymbol{h}}_{k}$
are generated according to i.i.d. complex Gaussian distributions with
zero mean and unit variance. The channel estimation error $\boldsymbol{e}_{k}$
also has i.i.d. complex Gaussian entries with zero mean and variance
$0.002$. The target average rate for all users is set to be the same
as $r_{k}=1$. The noise variance for all users is set to be 0.1.
Finally, the parameters $\rho^{t},\gamma^{t}$ are chosen as $\rho^{t}=\frac{1}{\left(1+n\right)^{0.9}}$,
$\gamma^{t}=\frac{15}{15+n}$. Similar step sizes have also been used
in the simulations in \cite{Yang_TSP2016_SSCA}. The specific values
for the coefficients such as $0.9$ and 15 are tuned to achieve a
good empirical convergence speed. 

\begin{figure}
\begin{centering}
\includegraphics[clip,width=85mm]{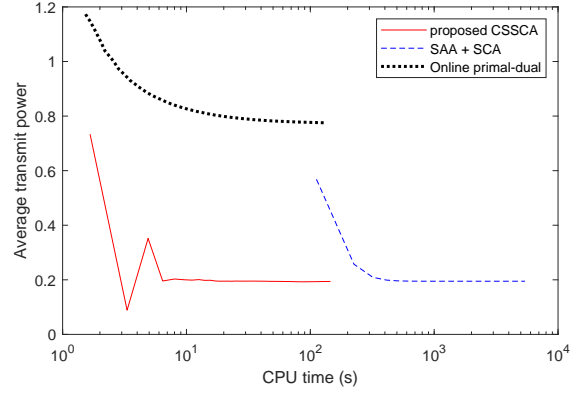}
\par\end{centering}
\caption{\label{fig:exm1a}Average transmit power versus the CPU time. {\small{}Simulation
software: Matlab R2018a. Simulation platform: Windows 10 x64 machine
with Intel i7-8550U CPU and 16 GB RAM. The same simulation platform
is used in Fig. 3 - 5.}}
\end{figure}

\begin{figure}
\begin{centering}
\includegraphics[clip,width=85mm]{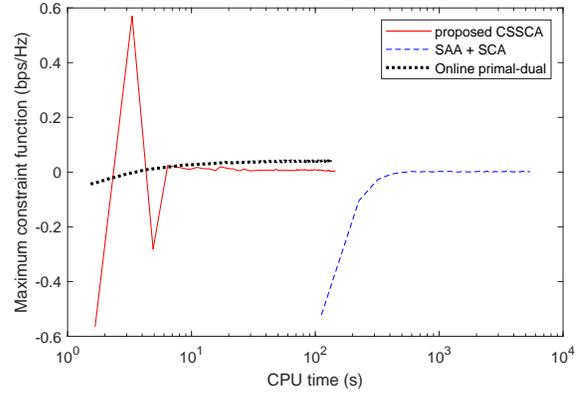}
\par\end{centering}
\caption{\label{fig:exm1b}Maximum constraint function versus the CPU time.}
\end{figure}

In Fig. \ref{fig:exm1a} and \ref{fig:exm1b}, we plot the objective
function (average transmit power) and maximum constraint function
(target average rate minus achieved average rate) versus the CPU time
respectively. The CSSCA and SAA converge to the same average transmit
power with all target average rates satisfied with high accuracy.
However, the online primal-dual algorithm cannot converge properly
and has much higher average transmit power. The CPU time required
to achieve a good convergence accuracy in the proposed CSSCA is much
less than that in the SAA. Although the CPU time depends on implementation
details, the codes for implementing CSSCA and SAA are very similar
except that SAA involves more number of system state samples at each
iteration. Therefore, the order-wise difference between the CPU times
of CSSCA and SAA is a strong evidence that the proposed CSSCA is more
efficient than SAA.

\subsection{Robust Beamforming Design}

The original robust beamforming design problem in (\ref{eq:RBFCcin})
is a chance constraint problem. To apply the proposed CSSCA, we first
approximate the step function using the smooth function in (\ref{eq:ustep}),
where a parameter $\theta$ is used to control the approximation error,
and then obtain a smooth approximation of (\ref{eq:RBFCcin}) in (\ref{eq:RBFCcin-1}).
Problem (\ref{eq:RBFCcin-1}) is an instance of (\ref{eq:mainP})
and the constraint can be written as $\mathbb{E}\left[g_{k}\left(\boldsymbol{w},\boldsymbol{H}\right)\right]\leq0,\forall k$
with $g_{k}\left(\boldsymbol{w},\boldsymbol{H}\right)=\hat{u}_{\theta}\left(s_{k}\left(\boldsymbol{w},\boldsymbol{H}\right)\right)$,
where
\begin{align*}
s_{k}\left(\boldsymbol{w},\boldsymbol{H}\right) & =\eta_{k}\left(\sum_{i\neq k}\left|\boldsymbol{h}_{k}^{H}\boldsymbol{w}_{i}\right|^{2}+\sigma_{k}^{2}\right)-\left|\boldsymbol{h}_{k}^{H}\boldsymbol{w}_{k}\right|^{2},
\end{align*}
and $\boldsymbol{w}=\left\{ \boldsymbol{w}_{i}\right\} _{i=1}^{K}$
is the set of all beamforming vectors. 

We choose to use the recursive surrogate function in (\ref{eq:TSF}),
but with multiple system state samples to generate the sample surrogate
function in (\ref{eq:ghead}) at each iteration. Specifically, we
first calculate the gradient of $g_{k}\left(\boldsymbol{w},\boldsymbol{H}\right)$
with respect to $\boldsymbol{w}_{i}$ as
\begin{equation}
\nabla_{\boldsymbol{w}_{i}}g_{k}\left(\boldsymbol{w},\boldsymbol{H}\right)=\begin{cases}
2\hat{u}_{\theta}^{'}\left(s_{k}\left(\boldsymbol{w},\boldsymbol{H}\right)\right)\eta_{k}\boldsymbol{h}_{k}^{H}\boldsymbol{w}_{i}\boldsymbol{h}_{k} & i\neq k\\
-2\hat{u}_{\theta}^{'}\left(s_{k}\left(\boldsymbol{w},\boldsymbol{H}\right)\right)\boldsymbol{h}_{k}^{H}\boldsymbol{w}_{i}\boldsymbol{h}_{k} & i=k
\end{cases}.\label{eq:gadexm2}
\end{equation}
Then we can obtain the expression of the recursive surrogate function
using (\ref{eq:TSF}), (\ref{eq:ghead}) and (\ref{eq:gadexm2}),
and implement the proposed CSSCA for Problem (\ref{eq:RBFCcin-1}).

As for the baseline algorithms, we use the Bernstein method proposed
in \cite{Wang_ICASSP11_Bernstein-type}. The Bernstein method usually
achieves an SINR outage probability that is less than the target and
thus is conservative. In the simulations, we also consider another
baseline which combines the Bernstein method with a bisection search
to further improve the performance. The details of this combined method
can be found in \cite{Wang_ICASSP11_Bernstein-type}.

\textit{Numerical Results:} We use a similar simulation configuration
as that in \cite{Wang_ICASSP11_Bernstein-type}. There are $n=3$
antennas and $K=3$ users. The SINR targets for all users are the
same: $\eta_{k}=5$ dB, $\forall k$. We set the value of the smooth
parameter $\theta=400$. The channel estimates $\left\{ \hat{\boldsymbol{h}}_{k}\right\} $
and channel estimation error $\left\{ \boldsymbol{e}_{k}\right\} $
have the same distributions as that in Example \ref{exa:Portfolio-optimization}.
The noise variances for all users are set to be 0.01. Finally, the
parameters $\rho^{t},\gamma^{t}$ are chosen as $\rho^{t}=\left(\frac{1}{1+n}\right)^{0.5}$,
$\gamma^{t}=\left(\frac{1}{1+n}\right)^{0.6}$.

\begin{table}
\centering{}{\small{}\caption{\label{tab:exm2}Comparison of the feasibility rate and average transmit
power.}
}{\footnotesize{}}%
\begin{tabular}{|l|l|l|l|}
\hline 
 & CSSCA & Bernstein & Combined\tabularnewline
\hline 
Feasibility rate & 94.33\%  & 94.02\% & 97.79\%\tabularnewline
Average power & 0.4877 & 1.8235 & 0.3341\tabularnewline
\hline 
\end{tabular}
\end{table}

In Table \ref{tab:exm2}, we examine the feasibility rates and the
average transmit power of the three algorithms. To this end, 5000
sets of channel estimates $\left\{ \hat{\boldsymbol{h}}_{k}\right\} $
were generated. It can be seen that CSSCA and Bernstein exhibit a
similar feasibility rate (a solution found by an algorithm is feasible
if it satisfies the SINR outage probability constraint in (\ref{eq:RBFCcin})
with finite transmit power), which is slightly smaller than that achieved
by the combined method. The combined method consumes the lowest transmit
power and the proposed CSSCA consumes a lower transmit power than
the Bernstein method. The proposed CSSCA works for any channel estimation
error distributions, while the Bernstein methods only work for Gaussian
error distributions.

\subsection{Massive MIMO Hybrid Beamforming Design}

In the massive MIMO hybrid beamforming design problem in (\ref{eq:NSP}),
the objective and constraint can be written as $\mathbb{E}\left[g_{0}\left(\boldsymbol{\Theta},\boldsymbol{p},\boldsymbol{H}\right)\right]$
and $\mathbb{E}\left[g_{1}\left(\boldsymbol{\Theta},\boldsymbol{p},\boldsymbol{H}\right)\right]\leq0$,
respectively, where $g_{0}\left(\boldsymbol{\Theta},\boldsymbol{p},\boldsymbol{H}\right)=\log\left(1+\frac{\left|\boldsymbol{h}_{k}^{H}\boldsymbol{F}\boldsymbol{g}_{k}\right|^{2}}{\sum_{i\neq k}\left|\boldsymbol{h}_{k}^{H}\boldsymbol{F}\boldsymbol{g}_{i}\right|^{2}+1}\right)$
and $g_{1}\left(\boldsymbol{\Theta},\boldsymbol{p},\boldsymbol{H}\right)=Tr\left(\boldsymbol{F}\boldsymbol{G}\boldsymbol{G}^{H}\boldsymbol{F}^{H}\right)-P$.
In the proposed CSSCA, we consider the following surrogate function
for the objective function:
\begin{align}
\bar{f}_{0}^{t}\left(\boldsymbol{\Theta},\boldsymbol{p}\right) & =f^{t}+Tr\left(\left(\mathbf{F}_{\Theta}^{t}\right)^{T}\left(\boldsymbol{\Theta}-\boldsymbol{\Theta}^{t}\right)\right)+\left(\mathbf{f}_{p}^{t}\right)^{T}\left(\boldsymbol{p}-\boldsymbol{p}^{t}\right),\nonumber \\
 & +\tau Tr\left(\left(\boldsymbol{\Theta}-\boldsymbol{\Theta}^{t}\right)\left(\boldsymbol{\Theta}-\boldsymbol{\Theta}^{t}\right)^{T}\right)+\tau\left\Vert \boldsymbol{p}-\boldsymbol{p}^{t}\right\Vert ^{2},\label{eq:sfexm3}
\end{align}
where $\mathbf{F}_{\Theta}^{t}$ and $\mathbf{f}_{p}^{t}$ can be
calculated recursively as 
\begin{align}
\mathbf{F}_{\Theta}^{t} & =\left(1-\rho^{t}\right)\mathbf{F}_{\Theta}^{t-1}+\rho^{t}\nabla_{\boldsymbol{\Theta}}g_{0}\left(\boldsymbol{\Theta}^{t},\boldsymbol{p}^{t},\boldsymbol{H}^{t}\right),\nonumber \\
\mathbf{f}_{p}^{t} & =\left(1-\rho^{t}\right)\mathbf{f}_{p}^{t-1}+\rho^{t}\nabla_{\boldsymbol{p}}g_{0}\left(\boldsymbol{\Theta}^{t},\boldsymbol{p}^{t},\boldsymbol{H}^{t}\right),\label{eq:frcexm3}
\end{align}
$\boldsymbol{H}^{t}$ is the channel sample obtained at the $t$-th
iteration, and the constant $f^{t}$ can be calculated as
\[
f^{t}=\frac{1}{t}\sum_{j=1}^{t}g_{0}\left(\boldsymbol{\Theta}^{t},\boldsymbol{p}^{t},\boldsymbol{H}^{j}\right).
\]
(\ref{eq:sfexm3}) is a special case of the structured surrogate function
in (\ref{eq:SSF}) with zero convex component $g_{0}^{c}\left(\boldsymbol{\Theta},\boldsymbol{p},\boldsymbol{H}\right)=0$.
The surrogate function $\bar{f}_{1}^{t}\left(\boldsymbol{\Theta},\boldsymbol{p}\right)$
for the constraint function is similar. 

The gradients of $g_{0}\left(\boldsymbol{\Theta},\boldsymbol{p},\boldsymbol{H}\right)$
w.r.t. $\boldsymbol{\Theta}$ and $\boldsymbol{p}$ in (\ref{eq:frcexm3})
are given by
\begin{align*}
\nabla_{\boldsymbol{\Theta}}g_{0}\left(\boldsymbol{\Theta},\boldsymbol{p},\boldsymbol{H}\right) & =\frac{\sum_{i}\boldsymbol{A}_{k,i}^{\theta}}{\varGamma_{k}}-\frac{\sum_{i\neq k}\boldsymbol{A}_{k,i}^{\theta}}{\varGamma_{-k}},\\
\nabla_{\boldsymbol{p}}g_{0}\left(\boldsymbol{\Theta},\boldsymbol{p},\boldsymbol{H}\right) & =\frac{\sum_{i}\boldsymbol{a}_{k,i}^{p}}{\varGamma_{k}}-\frac{\sum_{i\neq k}\boldsymbol{a}_{k,i}^{p}}{\varGamma_{-k}},
\end{align*}
where $\varGamma_{k}=\sum_{i}\left|\boldsymbol{h}_{k}^{H}\boldsymbol{F}\boldsymbol{g}_{i}\right|^{2}+1$,
$\varGamma_{-k}=\sum_{i\neq k}\left|\boldsymbol{h}_{k}^{H}\boldsymbol{F}\boldsymbol{g}_{i}\right|^{2}+1$,
\begin{align*}
\boldsymbol{A}_{k,i}^{\theta} & =\mathfrak{R}\left[j\boldsymbol{F}^{*}\circ2\left(\boldsymbol{H}_{F}^{H}\boldsymbol{A}_{i}\boldsymbol{H}_{F}\boldsymbol{F}-\boldsymbol{B}_{i}\boldsymbol{F}\right)\right],\\
\boldsymbol{a}_{k,i}^{p} & =Diag\left[\boldsymbol{h}_{k}\boldsymbol{h}_{k}^{H}\boldsymbol{H}_{F}\boldsymbol{F}\boldsymbol{F}^{H}\boldsymbol{h}_{k}\boldsymbol{h}_{k}^{H}\boldsymbol{F}\boldsymbol{F}^{H}\boldsymbol{H}_{F}^{H}\boldsymbol{I}_{i}\right],
\end{align*}
where $\circ$ denotes the Hadamard product, $Diag\left(\boldsymbol{M}\right)$
denotes a vector consisting of the diagonal elements of the matrix
$\boldsymbol{M}$, 
\begin{align*}
\boldsymbol{H}_{F} & =\left(\boldsymbol{H}\boldsymbol{F}\boldsymbol{F}^{H}\boldsymbol{H}^{H}+\frac{K}{P}\boldsymbol{I}\right)^{-1}\boldsymbol{H},\\
\boldsymbol{A}_{i} & =\mathfrak{S}\left[\boldsymbol{H}\boldsymbol{F}\boldsymbol{F}^{H}\boldsymbol{h}_{k}\boldsymbol{h}_{k}^{H}\boldsymbol{F}\boldsymbol{F}^{H}\boldsymbol{H}_{F}^{H}\boldsymbol{P}_{i}\right],\\
\boldsymbol{B}_{i} & =\mathfrak{S}\left[\boldsymbol{h}_{k}\boldsymbol{h}_{k}^{H}\boldsymbol{F}\boldsymbol{F}^{H}\boldsymbol{H}_{F}^{H}\boldsymbol{P}_{i}\boldsymbol{H}_{F}\right],
\end{align*}
$\mathfrak{S}\left[\boldsymbol{M}\right]\triangleq\boldsymbol{M}+\boldsymbol{M}^{H}$,
and $\boldsymbol{P}_{i}$ ($\boldsymbol{I}_{i}$) denotes a $K\times K$
matrix with $\left[\boldsymbol{P}_{i}\right]_{i,i}=p_{i}$ ($\left[\boldsymbol{I}_{i}\right]_{i,i}=1$)
and all other elements being zero. Similarly, the gradients of $g_{1}\left(\boldsymbol{\Theta},\boldsymbol{p},\boldsymbol{H}\right)$
w.r.t. $\boldsymbol{\Theta}$ and $\boldsymbol{p}$ (which are required
to construct the surrogate function $\bar{f}_{1}^{t}\left(\boldsymbol{\Theta},\boldsymbol{p}\right)$
of the constraint) are given by
\begin{align*}
\nabla_{\boldsymbol{\Theta}}g_{1}\left(\boldsymbol{\Theta},\boldsymbol{p},\boldsymbol{H}\right) & =\mathfrak{R}\left[j\boldsymbol{F}^{*}\circ\left(2\boldsymbol{H}_{F}^{H}\boldsymbol{A}\boldsymbol{H}_{F}\boldsymbol{F}-2\boldsymbol{B}\boldsymbol{F}\right)\right],\\
\nabla_{\boldsymbol{p}}g_{1}\left(\boldsymbol{\Theta},\boldsymbol{p},\boldsymbol{H}\right) & =Diag\left[\boldsymbol{H}_{F}\boldsymbol{F}\boldsymbol{F}^{H}\boldsymbol{F}\boldsymbol{F}^{H}\boldsymbol{H}_{F}^{H}\right],
\end{align*}
where
\begin{align*}
\boldsymbol{A} & =\mathfrak{S}\left[\boldsymbol{H}\boldsymbol{F}\boldsymbol{F}^{H}\boldsymbol{F}\boldsymbol{F}^{H}\boldsymbol{H}_{F}^{H}\boldsymbol{P}\right],\\
\boldsymbol{B} & =\mathfrak{S}\left[\boldsymbol{F}\boldsymbol{F}^{H}\boldsymbol{H}_{F}^{H}\boldsymbol{P}\boldsymbol{H}_{F}\right].
\end{align*}

With the surrogate function in (\ref{eq:sfexm3}), the feasible update
in (\ref{eq:Pitert-1}) is a quadratic programming with a closed-form
solution. On the other hand, the objective update in (\ref{eq:Pitert})
is a simple optimization problem with a quadratic objective function
and a quadratic constraint, which can be easily solved by the Lagrange
dual method. Specifically, for given Lagrange multiplier, the optimal
primal variable that maximizes the Lagrange function has a closed-form
solution. Then we can use a bisection method to find the optimal Lagrange
multiplier. The details are omitted for conciseness. 

We consider the SAA with $N=200$ channel samples as the baseline
algorithm and the resulting deterministic optimization problem has
a non-convex constraint, which is again solved using the deterministic
SCA method in \cite{Meisam_thesis14_BSUM}. The SCA method uses a
surrogate function which has similar form as that in (\ref{eq:sfexm3}).
The online primal-dual algorithm is also included as a baseline. The
same Lagrange dual method is used to solve the convex subproblem in
each iteration of the ``SAA + SCA'' baseline.

\begin{figure}
\begin{centering}
\includegraphics[clip,width=85mm]{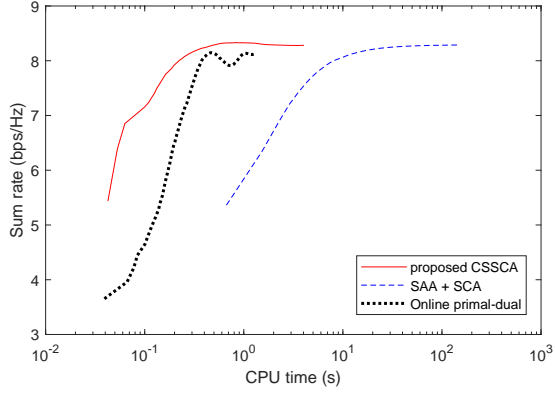}
\par\end{centering}
\caption{\label{fig:exm3a}Sum rate versus the CPU time.}
\end{figure}

\begin{figure}
\begin{centering}
\includegraphics[clip,width=85mm]{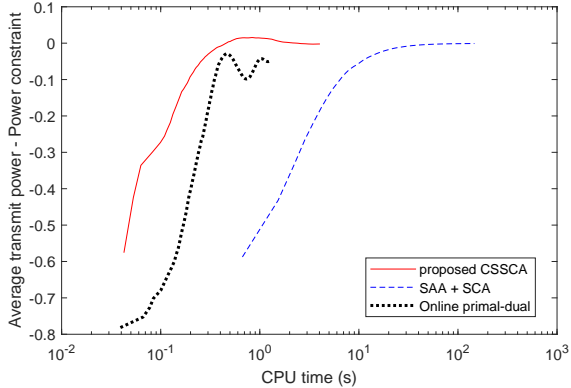}
\par\end{centering}
\caption{\label{fig:exm3b}Average transmit power minus power constraint $P$
versus the CPU time.}
\end{figure}

\textit{Numerical Results:} In the simulations, the massive MIMO BS
is equipped with $M=64$ antennas and $S=8$ transmit RF chains. There
are $K=4$ users. We consider a spatially correlated channel model:
$\boldsymbol{H}=\boldsymbol{R}^{1/2}\boldsymbol{H}_{w}$, where $\boldsymbol{H}_{w}$
has i.i.d. complex Gaussian entries with zero mean and unit variance
and $\boldsymbol{R}$ is the spatial correlation matrix. Since the
massive MIMO channel is usually highly correlated \cite{Liu_TSP14_RFprecoding},
we assume that $\boldsymbol{R}$ is rank deficient. Specifically,
we let $\boldsymbol{R}=8\boldsymbol{U}\boldsymbol{U}^{H}$, where
$\boldsymbol{U}^{64\times8}$ is a randomly generated semi-unitary
matrix and the coefficient $8$ is chosen to normalize $\boldsymbol{R}$
such that $Tr\left(\boldsymbol{R}\right)=M$. The power constraint
is set to be $P=0$ dB. Finally, the parameters $\rho^{t},\gamma^{t}$
are chosen as $\rho^{t}=\frac{1}{\left(1+n\right)^{2/3}}$, $\gamma^{t}=\frac{2}{2+n}$.

In Fig. \ref{fig:exm3a} and \ref{fig:exm3b}, we plot the objective
function (sum rate) and constraint function (average transmit power
minus power constraint $P$) versus the CPU time respectively. The
CSSCA and SAA converge to the same sum rate with the average power
constraint satisfied with high accuracy. However, the online primal-dual
algorithm converges very slowly and achieves a lower sum rate. Note
that although different implementations are used to solve the per-iteration
convex subproblems in Fig. \ref{fig:exm1a}, \ref{fig:exm1b} and
Fig. \ref{fig:exm3a}, \ref{fig:exm3b}, we can see similar order-wise
differences between the CPU times of CSSCA and SAA in all these figures.
This strongly suggests that the proposed CSSCA is much more efficient
than SAA. Moreover, since SAA is an offline method, it requires a
\textit{channel sample collection phase} to obtain a sufficiently
large number of channel samples before calculating the optimized RF
precoder. As a result, the performance will be bad at the channel
sample collection phase, which may last for a few hundreds channel
coherence intervals. On the other hand, the proposed CSSCA is an online
method which can update the RF precoder whenever a new channel sample
is obtained. As a result, it can achieve a better overall performance
compared to the SAA.

\section{Conclusions\label{sec:Conclusion}}

We consider a general stochastic optimization problem where both objective
and constraint functions are non-convex and involve expectations over
random states. We propose a CSSCA algorithm to find a stationary point
of the problem. At each iteration, the algorithm first updates the
convex surrogate functions for the objective and constraints based
on the observed random state and current iterate. If the convex approximation
problem constructed from the surrogate functions is feasible, the
algorithm performs an objective update by solving the convex approximation
problem. Otherwise, it performs a feasibility update by minimizing
the maximum of the surrogate functions for constraints. We show that
under some technical conditions, the algorithm converges to a stationary
point of the original problem almost surely. We also gives a parallel
implementation for the algorithm when the constraint function is decoupled.
The parallel version of the CSSCA is desirable for solving large-scale
stochastic optimization problems such as those that rise in machine
learning and big data. Finally, we use several important application
examples to illustrate the effectiveness of the proposed algorithm. 

\appendix

\subsection{Proof of Proposition \ref{prop:Validity-of-theRSF}\label{subsec:Proof-of-Proposition}}

Assumption \ref{asm:convAnew}-1 and Assumption \ref{asm:convAnew}-3
follow immediately from Assumption \ref{asm:convRSF}. The rest of
the proof relies on (\cite{Ruszczyski_MP80_SPthem}, Lemma 1), which
is restated below for completeness.
\begin{lem}
\label{lem:refkeu}Let $\left(\Omega,\mathcal{F},\mathbb{P}\right)$
be a probability space and let $\left\{ \mathcal{F}_{t}\right\} $
be an increasing sequence of $\sigma$-field contained in $\mathcal{F}$.
Let $\left\{ \boldsymbol{\eta}^{t}\right\} ,\left\{ \boldsymbol{z}^{t}\right\} $
be sequences of $\mathcal{F}_{t}$-measurable random vectors satisfying
the relations 
\begin{align*}
\boldsymbol{z}^{t+1} & =\Pi_{\mathcal{Z}}\left(\boldsymbol{z}^{t}+\rho^{t}\left(\boldsymbol{\zeta}^{t}-\boldsymbol{z}^{t}\right)\right),\boldsymbol{z}^{0}\in\mathcal{Z},\\
\mathbb{E}\left[\boldsymbol{\zeta}^{t}|\mathcal{F}_{t}\right] & =\boldsymbol{\eta}^{t}+\boldsymbol{b}^{t},
\end{align*}
where $\rho^{t}\geq0$ and the set $\mathcal{Z}$ is convex and closed,
$\Pi_{\mathcal{Z}}\left(\cdot\right)$ denotes projection on $\mathcal{Z}$.
Next, let

(a) all accumulation points of the sequence $\left\{ \boldsymbol{\eta}^{t}\right\} $
belong to $\mathcal{Z}$ w.p.1.,

(b) there exists a constant $C$ such that $\mathbb{E}\left[\left\Vert \boldsymbol{\zeta}^{t}\right\Vert ^{2}|\mathcal{F}_{t}\right]\leq C$
for all $t\geq0$,

(c) $\sum_{t=0}^{\infty}\mathbb{E}\left[\left(\rho^{t}\right)^{2}+\rho^{t}\left\Vert \boldsymbol{b}^{t}\right\Vert \right]<\infty$,

(d) $\sum_{t=0}^{\infty}\rho^{t}=\infty$, and (e) $\left\Vert \boldsymbol{\eta}^{t+1}-\boldsymbol{\eta}^{t}\right\Vert /\rho^{t}\rightarrow0$
w.p.1.

Then $\boldsymbol{z}^{t}-\boldsymbol{\eta}^{t}\rightarrow0$ w.p.1.
\end{lem}

Using this result, we can prove the following key lemma.
\begin{lem}
\label{lem:fconv}Under Assumption \ref{asm:convP}, \ref{asm:convRSF}
and \ref{asm:convS}, we have
\begin{align*}
\lim_{t\rightarrow\infty}\left|\bar{f}_{i}^{t}\left(\boldsymbol{x}^{t}\right)-f_{i}\left(\boldsymbol{x}^{t}\right)\right| & =0,\\
\lim_{t\rightarrow\infty}\left\Vert \nabla\bar{f}_{i}^{t}\left(\boldsymbol{x}^{t}\right)-\nabla f_{i}\left(\boldsymbol{x}^{t}\right)\right\Vert  & =0,\\
\lim_{t\rightarrow\infty}\left|\bar{f}_{i}^{t}\left(\boldsymbol{x}\right)-\bar{g}_{i}\left(\boldsymbol{x},\boldsymbol{x}^{t}\right)\right| & =0,\forall\boldsymbol{x}\in\mathcal{X},
\end{align*}
for $i=0,...,m$\textup{ w.p.1.}, where $\bar{g}_{i}\left(\boldsymbol{x},\boldsymbol{x}^{t}\right)\triangleq\mathbb{E}\left[\hat{g}_{i}\left(\boldsymbol{x},\boldsymbol{x}^{t},\boldsymbol{\xi}\right)\right]$.
\end{lem}

\begin{IEEEproof}
Lemma \ref{lem:fconv} is a consequence of Lemma \ref{lem:refkeu}.
We only need to verify that all the technical conditions therein are
satisfied by the problem in Lemma \ref{lem:fconv} and the proof is
similar to that of (\cite{Yang_TSP2016_SSCA}, Lemma 1). The details
are omitted for conciseness.
\end{IEEEproof}

Assumption \ref{asm:convBnew} follows immediately from Lemma \ref{lem:fconv}.
To prove Assumption \ref{asm:convAnew}-2, it follows from Lemma \ref{lem:fconv}
that 
\begin{equation}
\bar{f}_{i}^{t}\left(\boldsymbol{x}\right)=\bar{g}_{i}\left(\boldsymbol{x},\boldsymbol{x}^{t}\right)+e_{i}\left(t\right),\label{eq:fibconv}
\end{equation}
where $\lim_{t\rightarrow\infty}e_{i}\left(t\right)\rightarrow0$.
From Assumption \ref{asm:convRSF}, $\bar{g}_{i}\left(\boldsymbol{x},\boldsymbol{x}^{t}\right)$
is Lipschitz continuous in $\boldsymbol{x}^{t}$ and thus
\begin{equation}
\left|\bar{g}_{i}\left(\boldsymbol{x},\boldsymbol{x}^{t_{1}}\right)-\bar{g}_{i}\left(\boldsymbol{x},\boldsymbol{x}^{t_{2}}\right)\right|\leq B\left\Vert \boldsymbol{x}^{t_{1}}-\boldsymbol{x}^{t_{2}}\right\Vert ,\label{eq:LCfibar}
\end{equation}
for some constant $B>0$. Combining (\ref{eq:fibconv}) and (\ref{eq:LCfibar}),
we have
\[
\bar{f}_{i}^{t_{1}}\left(\boldsymbol{x}\right)-\bar{f}_{i}^{t_{2}}\left(\boldsymbol{x}\right)\leq B\left\Vert \boldsymbol{x}^{t_{1}}-\boldsymbol{x}^{t_{2}}\right\Vert +e_{g}\left(t_{1},t_{2}\right),
\]
 where $\lim_{t_{1},t_{2}\rightarrow\infty}e_{g}\left(t_{1},t_{2}\right)=0$,
from which Assumption \ref{asm:convAnew}-2 follows.

\subsection{Proof of Lemma \ref{lem:Convergence-surrogate}\label{subsec:Proof-of-Lemmaconvf}}

Due to Assumption \ref{asm:convAnew}, the families of functions $\left\{ \bar{f}_{i}^{t_{j}}\left(\boldsymbol{x}\right)\right\} $
are equicontinuous. Moreover, they are bounded and defined over a
compact set $\mathcal{X}$. Hence the Arzela\textendash Ascoli theorem
\cite{Dunford_Inpub58_LO} implies that, by restricting to a subsequence,
there exists uniformly continuous functions $\hat{f}_{i}\left(\boldsymbol{x}\right)$
such that (\ref{eq:ghfhead}) is satisfied. Finally, (\ref{eq:convsur})
follows immediately from (\ref{eq:ghfhead}) and Lemma \ref{lem:fconv}.

\subsection{Proof of Theorem \ref{thm:Convergence-of-Algorithm1-1} \label{subsec:Proof-of-Theorem-local}}

1. We first give a lemma that is crucial for the convergence proof. 
\begin{lem}
\label{lem:keylem}Suppose Assumptions \ref{asm:convP}, \ref{asm:convAnew}
and \ref{asm:convBnew} are satisfied. Moreover, suppose $\mathcal{X}_{A}^{*}\cap\overline{\mathcal{X}}_{C}^{*}=\emptyset$,
where $\mathcal{X}_{A}^{*}$ is the set of limiting points of Algorithm
1. Let $\left\{ \boldsymbol{x}^{t}\right\} _{t=1}^{\infty}$ denote
the sequence of iterates generated by Algorithm 1. We have 
\begin{align*}
\limsup_{t\rightarrow\infty}\max_{i\in\left\{ 1,...,m\right\} }f_{i}\left(\boldsymbol{x}^{t}\right) & \leq0,\text{ w.p.1.}\\
\lim_{t\rightarrow\infty}\left\Vert \bar{\boldsymbol{x}}^{t}-\boldsymbol{x}^{t}\right\Vert  & =0,\text{ w.p.1.}
\end{align*}

\end{lem}
The lemma states that when $\mathcal{X}_{A}^{*}\cap\overline{\mathcal{X}}_{C}^{*}=\emptyset$,
the algorithm will converge to the feasible region, and the gap between
$\bar{\boldsymbol{x}}^{t}$ and $\boldsymbol{x}^{t}$ converges to
zero, almost surely. Please refer to Appendix \ref{subsec:Proof-of-keyLemma}
for the proof.

2. Then we prove that under the conditions in Theorem \ref{thm:Convergence-of-Algorithm1-1},
we have $\boldsymbol{x}^{t}\notin\overline{\mathcal{X}}_{C}^{*},\forall t$
and thus $\mathcal{X}_{A}^{*}\cap\overline{\mathcal{X}}_{C}^{*}=\emptyset$
holds true with probability 1.

When $\overline{\mathcal{X}}_{C}^{*}=\emptyset$, $\boldsymbol{x}^{t}\notin\overline{\mathcal{X}}_{C}^{*},\forall t$
is automatically satisfied. Therefore, we shall focus on the non-trivial
case when $\overline{\mathcal{X}}_{C}^{*}\neq\emptyset$. Let
\[
\mathcal{L}\left(\alpha\right)=\left\{ \boldsymbol{x}:\:f\left(\boldsymbol{x}\right)\leq\alpha\right\} 
\]
denote a sublevel set of $f\left(\boldsymbol{x}\right)$ at level
$\alpha$. Let $\alpha_{C}=\min_{\boldsymbol{x}\in\overline{\mathcal{X}}_{C}^{*}}\:f\left(\boldsymbol{x}\right)$.
By the definition of $\overline{\mathcal{X}}_{C}^{*}$, we must have
$\alpha_{C}>0$. Since $f\left(\boldsymbol{x}^{0}\right)\leq0$, we
must have $\boldsymbol{x}^{0}\in\mathcal{L}\left(0.5\alpha_{C}\right)$.
Let $\mathcal{X}_{S}$ be a compact subset of $\mathcal{L}\left(0.5\alpha_{C}\right)$
such that all the points in $\mathcal{X}_{S}$ is connected with $\boldsymbol{x}^{0}$.
Note that by definition, $\mathcal{X}_{S}\cap\overline{\mathcal{X}}_{C}^{*}=\emptyset$.
Let $\hat{\alpha}=0.25\alpha_{C}$ and $\mathcal{L}\left(\hat{\alpha},\mathcal{X}_{S}\right)=\mathcal{L}\left(\hat{\alpha}\right)\cap\mathcal{X}_{S}$.
Since $f\left(\boldsymbol{x}\right)$ is Lipschitz continuous, there
exists a constant $L>0$ such that 
\begin{equation}
\min_{\boldsymbol{x}\in\partial\mathcal{L}\left(\hat{\alpha},\mathcal{X}_{S}\right)}\left\Vert \boldsymbol{x}-\boldsymbol{x}^{0}\right\Vert \geq L\left(0.25\alpha_{C}-f\left(\boldsymbol{x}^{0}\right)\right)\geq0.25L\alpha_{C},\label{xxp-1}
\end{equation}
where $\partial\mathcal{S}$ denote the boundary of a set $\mathcal{S}$.

By redefine the set $\mathcal{T}_{\epsilon},\mathcal{T}_{\epsilon}^{'}$
in Appendix \ref{subsec:Proof-of-keyLemma} as $\mathcal{T}_{\epsilon}=\left\{ t:\:f\left(\boldsymbol{x}^{t}\right)\geq\epsilon,\boldsymbol{x}^{t}\in\mathcal{X}_{S}\right\} $,
$\mathcal{T}_{\epsilon}^{'}=\mathcal{T}_{\epsilon}\cap\left\{ t\geq t_{\epsilon}\right\} $,
and following the same analysis as in Appendix \ref{subsec:Proof-of-keyLemma},
it can be shown that (\ref{eq:gapf}) and (\ref{eq:gapf1}) still
hold since $\mathcal{X}_{S}\cap\overline{\mathcal{X}}_{C}^{*}=\emptyset$.
Suppose we choose $\gamma^{0}<0.25L\alpha_{C}/\left(R_{\mathcal{X}}t_{\epsilon}\right)$,
where $R_{\mathcal{X}}\triangleq\max_{\boldsymbol{x},\boldsymbol{y}\in\mathcal{X}}\left\Vert \boldsymbol{x}-\boldsymbol{y}\right\Vert $
is the diameter of $\mathcal{X}$. Then from (\ref{eq:updatext})
and (\ref{xxp-1}), we must have $\boldsymbol{x}^{t}\in\mathcal{L}\left(\hat{\alpha},\mathcal{X}_{S}\right)$
for $t\leq t_{\epsilon}$.

From (\ref{eq:gapf}), we know that $f\left(\boldsymbol{x}^{t}\right)$
will be decreased (almost surely) whenever $f\left(\boldsymbol{x}^{t}\right)\geq\epsilon$,
$t\geq t_{\epsilon}$ and $\boldsymbol{x}^{t}\in\mathcal{X}_{S}$.
Moreover, from the Lipschitz continuity $f\left(\boldsymbol{x}\right)$,
we have
\begin{equation}
\min_{\boldsymbol{x}\in\mathcal{L}\left(\hat{\alpha},\mathcal{X}_{S}\right),\boldsymbol{x}^{'}\in\partial\mathcal{X}_{S}}\left\Vert \boldsymbol{x}-\boldsymbol{x}^{'}\right\Vert \geq0.25L\alpha_{C}.\label{xxp}
\end{equation}
Since $\alpha_{C}>0$, we can always choose a sufficiently small $\epsilon$
such that $f\left(\boldsymbol{x}\right)>3\epsilon,\forall\boldsymbol{x}\in\partial\mathcal{X}_{S}$,
$\hat{\alpha}>2\epsilon$ and $0.25L\alpha_{C}>2\epsilon$. From (\ref{xxp}),
once $\boldsymbol{x}^{t}\in\mathcal{L}\left(\hat{\alpha},\mathcal{X}_{S}\right)$
for $t\geq t_{\epsilon}$, $\boldsymbol{x}^{t+1}$ must also belong
to $\mathcal{X}_{S}$ because $\left\Vert \boldsymbol{x}^{t+1}-\boldsymbol{x}^{t}\right\Vert \leq O(\gamma^{t})<\epsilon$
for sufficiently large $t_{\epsilon}$, and there are two cases. 

Case 1: $f\left(\boldsymbol{x}^{t}\right)\geq\epsilon$. In this case,
we have $f\left(\boldsymbol{x}^{t+1}\right)<f\left(\boldsymbol{x}^{t}\right)$
according to (\ref{eq:gapf}) and thus $\boldsymbol{x}^{t+1}\in\mathcal{L}\left(\hat{\alpha},\mathcal{X}_{S}\right)$
according to the definition of sublevel set, with probability 1. 

Case 2: $f\left(\boldsymbol{x}^{t}\right)<\epsilon$. From (\ref{eq:gapf1}),
we have $f\left(\boldsymbol{x}^{t+1}\right)<2\epsilon$ and thus $\boldsymbol{x}^{t+1}\in\mathcal{L}\left(\hat{\alpha},\mathcal{X}_{S}\right)$,
with probability 1. 

In any case, we have $\boldsymbol{x}^{t+1}\in\mathcal{L}\left(\hat{\alpha},\mathcal{X}_{S}\right)$
with probability 1. Therefore, once $\boldsymbol{x}^{t}\in\mathcal{L}\left(\hat{\alpha},\mathcal{X}_{S}\right)$
for $t\geq t_{\epsilon}$, it remains in $\mathcal{L}\left(\hat{\alpha},\mathcal{X}_{S}\right)$
with probability 1. Together with the fact that $\boldsymbol{x}^{t}\in\mathcal{L}\left(\hat{\alpha},\mathcal{X}_{S}\right),\forall t\leq t_{\epsilon}$,
we conclude that $\boldsymbol{x}^{t}\in\mathcal{L}\left(\hat{\alpha},\mathcal{X}_{S}\right)\subset\mathcal{X}_{S},\forall t$
with probability 1. Since $\mathcal{X}_{S}\cap\overline{\mathcal{X}}_{C}^{*}=\emptyset$,
we have $\boldsymbol{x}^{t}\notin\overline{\mathcal{X}}_{C}^{*},\forall t$
with probability 1.

3. Finally, we prove Theorem \ref{thm:Convergence-of-Algorithm1-1}.

Let $\left\{ \boldsymbol{x}^{t_{j}}\right\} _{j=1}^{\infty}$ denote
any subsequence converging to a limit point $\boldsymbol{x}^{*}$
that satisfies the Slater condition. Since $\boldsymbol{x}^{t}\notin\overline{\mathcal{X}}_{C}^{*},\forall t$
and $\mathcal{X}_{A}^{*}\cap\overline{\mathcal{X}}_{C}^{*}=\emptyset$
w.p.1., it follows from Lemma \ref{lem:keylem} (and its proof in
Appendix \ref{subsec:Proof-of-keyLemma}) that 
\begin{equation}
\lim_{j\rightarrow\infty}\left\Vert \bar{\boldsymbol{x}}^{t_{j}}-\boldsymbol{x}^{t_{j}}\right\Vert =0,\text{ w.p.1.},\label{eq:lem2rst}
\end{equation}
and
\begin{align}
\bar{\boldsymbol{x}}^{t_{j}}=\underset{\boldsymbol{x}\in\mathcal{X}}{\text{argmin}}\: & \bar{f}_{0}^{t_{j}}\left(\boldsymbol{x}\right)\label{eq:Pixbart-1}\\
s.t.\: & \bar{f}_{i}^{t_{j}}\left(\boldsymbol{x}\right)\leq\alpha^{t_{j}},i=1,....,m,\nonumber 
\end{align}
where 
\begin{equation}
\lim_{j\rightarrow\infty}\alpha^{t_{j}}=0,\text{ w.p.1.}\label{eq:alpha0}
\end{equation}
Moreover, from Lemma \ref{lem:Convergence-surrogate}, we have
\begin{align}
\lim_{j\rightarrow\infty}\bar{f}_{i}^{t_{j}}\left(\boldsymbol{x}\right) & =\hat{f}_{i}\left(\boldsymbol{x}\right),\:\forall\boldsymbol{x}\in\mathcal{X},\label{eq:ghfhead-1}
\end{align}
almost surely. Letting $j\rightarrow\infty$ in (\ref{eq:Pixbart-1}),
using (\ref{eq:lem2rst}), (\ref{eq:alpha0}), (\ref{eq:ghfhead-1})
and the Lipschitz continuity and strong convexity of $\bar{f}_{i}^{t}\left(\boldsymbol{x}\right),\hat{f}_{i}\left(\boldsymbol{x}\right),\forall i$,
we have
\begin{align}
\boldsymbol{x}^{*}=\underset{\boldsymbol{x}\in\mathcal{X}}{\text{argmin}}\: & \hat{f}_{0}\left(\boldsymbol{x}\right)\label{eq:Piterthead}\\
s.t.\: & \hat{f}_{i}\left(\boldsymbol{x}\right)\leq0,i=1,....,m.\nonumber 
\end{align}
Since the Slater condition is satisfied, the KKT condition of the
problem (\ref{eq:Piterthead}) implies that there exist $\lambda_{1},...,\lambda_{m}$
such that
\begin{align}
\nabla\hat{f}_{0}\left(\boldsymbol{x}^{*}\right)+\sum_{i}\lambda_{i}\nabla\hat{f}_{i}\left(\boldsymbol{x}^{*}\right) & =\boldsymbol{0}\nonumber \\
\hat{f}_{i}\left(\boldsymbol{x}^{*}\right) & \leq0,\:\forall i=1,...,m\nonumber \\
\lambda_{i}\hat{f}_{i}\left(\boldsymbol{x}^{*}\right) & =0,\:\forall i=1,...,m.\label{KKTPhead}
\end{align}
Finally, it follows from Lemma \ref{lem:Convergence-surrogate} and
(\ref{KKTPhead}) that $\boldsymbol{x}^{*}$ also satisfies the KKT
condition of Problem (\ref{eq:mainP}). This completes the proof.

\subsection{Proof of Lemma \ref{lem:keylem}\label{subsec:Proof-of-keyLemma}}

1. We first prove $\limsup_{t\rightarrow\infty}f\left(\boldsymbol{x}^{t}\right)\leq0$
 w.p.1., where $f\left(\boldsymbol{x}\right)=\max_{i\in\left\{ 1,...,m\right\} }f_{i}\left(\boldsymbol{x}\right)$. 

Let $\mathcal{T}_{\epsilon}=\left\{ t:\:f\left(\boldsymbol{x}^{t}\right)\geq\epsilon\right\} $
for any $\epsilon>0$. We show that $\mathcal{T}_{\epsilon}$ is a
finite set by contradiction. 

Suppose $\mathcal{T}_{\epsilon}$ is infinite. We first show that
$\liminf_{t\in\mathcal{T}_{\epsilon},t\rightarrow\infty}\left\Vert \bar{\boldsymbol{x}}^{t}-\boldsymbol{x}^{t}\right\Vert >0$
by contradiction. Suppose $\liminf_{t\in\mathcal{T}_{\epsilon},t\rightarrow\infty}\left\Vert \bar{\boldsymbol{x}}^{t}-\boldsymbol{x}^{t}\right\Vert =0$.
Then there exists a subsequence $t^{j}\in\mathcal{T}_{\epsilon}$
such that $\lim_{j\rightarrow\infty}\left\Vert \bar{\boldsymbol{x}}^{t_{j}}-\boldsymbol{x}^{t_{j}}\right\Vert =0$.
Let $\boldsymbol{x}^{\circ}$ denote a limiting point of the subsequence
$\left\{ \boldsymbol{x}^{t_{j}}\right\} $, and let $\hat{f}_{i}\left(\boldsymbol{x}\right),\forall i$
be the converged surrogate functions as defined in Lemma \ref{lem:Convergence-surrogate}.
According to the update rule of Algorithm 1, there are two cases.

Case 1: $\boldsymbol{x}^{\circ}$ is the optimal solution of the following
convex optimization problem: 

\begin{align}
\underset{\boldsymbol{x}\in\mathcal{X}}{\text{min}}\: & \hat{f}_{0}\left(\boldsymbol{x}\right)\label{eq:Pcase1}\\
s.t.\: & \hat{f}_{i}\left(\boldsymbol{x}\right)\leq0,i=1,....,m.\nonumber 
\end{align}
In this case, we have $f\left(\boldsymbol{x}^{\circ}\right)=\max_{i\in\left\{ 1,...,m\right\} }\hat{f}_{i}\left(\boldsymbol{x}^{\circ}\right)\leq0$,
which contradicts the definition of $\mathcal{T}_{\epsilon}$.

Case 2: $\boldsymbol{x}^{\circ}$ is the optimal solution of the following
convex optimization problem: 

\begin{align}
\underset{\boldsymbol{x}\in\mathcal{X},\alpha}{\text{min}}\: & \alpha\label{eq:Piterthead-1}\\
s.t.\: & \hat{f}_{i}\left(\boldsymbol{x}\right)\leq\alpha,i=1,....,m.\nonumber 
\end{align}
Since the Slater condition is satisfied (by choosing a sufficiently
large $\alpha$, we can always find a point $\boldsymbol{x}\in\mathcal{X}$
such that $\hat{f}_{i}\left(\boldsymbol{x}\right)<\alpha,i=1,....,m$),
the KKT condition of the problem (\ref{eq:Piterthead-1}) implies
that there exist $\lambda_{1},...,\lambda_{m}$ such that
\begin{align}
\sum_{i}\lambda_{i}\nabla\hat{f}_{i}\left(\boldsymbol{x}^{\circ}\right) & =\boldsymbol{0}\nonumber \\
1-\sum_{i}\lambda_{i} & =0\nonumber \\
\hat{f}_{i}\left(\boldsymbol{x}^{\circ}\right) & \leq\alpha,\:\forall i=1,...,m\nonumber \\
\lambda_{i}\left(\hat{f}_{i}\left(\boldsymbol{x}^{\circ}\right)-\alpha\right) & =0,\:\forall i=1,...,m.\label{KKTPhead1}
\end{align}
It follows from Lemma \ref{lem:Convergence-surrogate} and (\ref{KKTPhead1})
that $\boldsymbol{x}^{\circ}$ also satisfies the KKT condition of
Problem (\ref{eq:FP-2}). From the condition $\mathcal{X}_{A}^{*}\cap\overline{\mathcal{X}}_{C}^{*}=\emptyset$,
we have $f_{i}\left(\boldsymbol{x}^{\circ}\right)\leq0,i=1,...,m$,
which again contradicts the definition of $\mathcal{T}_{\epsilon}$.

Therefore, $\liminf_{t\in\mathcal{T}_{\epsilon},t\rightarrow\infty}\left\Vert \bar{\boldsymbol{x}}^{t}-\boldsymbol{x}^{t}\right\Vert >0$,
i.e., there exists a sufficiently large $t_{\epsilon}$ such that
\begin{equation}
\left\Vert \bar{\boldsymbol{x}}^{t}-\boldsymbol{x}^{t}\right\Vert \geq\epsilon^{'},\forall t\in\mathcal{T}_{\epsilon}^{'}\label{eq:gapxt}
\end{equation}
where $\epsilon^{'}>0$ is some constant and $\mathcal{T}_{\epsilon}^{'}=\mathcal{T}_{\epsilon}\cap\left\{ t\geq t_{\epsilon}\right\} $. 

Define function $\bar{f}^{t}\left(\boldsymbol{x}\right)=\max_{i\in\left\{ 1,...,m\right\} }\bar{f}_{i}^{t}\left(\boldsymbol{x}\right)$.
From Assumption \ref{asm:convAnew}, $\bar{f}_{i}^{t}\left(\boldsymbol{x}^{t}\right)$
is strongly convex, and thus
\begin{equation}
\nabla^{T}\bar{f}_{i}^{t}\left(\boldsymbol{x}^{t}\right)\boldsymbol{d}^{t}\leq-\eta\left\Vert \boldsymbol{d}^{t}\right\Vert ^{2}+\bar{f}_{i}^{t}\left(\bar{\boldsymbol{x}}^{t}\right)-\bar{f}_{i}^{t}\left(\boldsymbol{x}^{t}\right),\label{eq:ftdbound}
\end{equation}
where $\boldsymbol{d}^{t}=\bar{\boldsymbol{x}}^{t}-\boldsymbol{x}^{t}$,
and $\eta>0$ is some constant. From Assumption \ref{asm:convP},
the gradient of $f_{i}\left(\boldsymbol{x}\right)$ is Lipschitz continuous,
and thus there exists $L_{f}>0$ such that
\begin{align}
f_{i}\left(\boldsymbol{x}^{t+1}\right) & \leq f_{i}\left(\boldsymbol{x}^{t}\right)+\gamma^{t}\nabla^{T}f_{i}\left(\boldsymbol{x}^{t}\right)\boldsymbol{d}^{t}+L_{f}\left(\gamma^{t}\right)^{2}\left\Vert \boldsymbol{d}^{t}\right\Vert ^{2}\nonumber \\
 & =f\left(\boldsymbol{x}^{t}\right)+L_{f}\left(\gamma^{t}\right)^{2}\left\Vert \boldsymbol{d}^{t}\right\Vert ^{2}+f_{i}\left(\boldsymbol{x}^{t}\right)-f\left(\boldsymbol{x}^{t}\right)\nonumber \\
 & +\gamma^{t}\left(\nabla^{T}\bar{f}_{i}^{t}\left(\boldsymbol{x}^{t}\right)+\nabla^{T}f_{i}\left(\boldsymbol{x}^{t}\right)-\nabla^{T}\bar{f}_{i}^{t}\left(\boldsymbol{x}^{t}\right)\right)\boldsymbol{d}^{t}\nonumber \\
 & \overset{\textrm{a}}{\leq}f\left(\boldsymbol{x}^{t}\right)+f_{i}\left(\boldsymbol{x}^{t}\right)-f\left(\boldsymbol{x}^{t}\right)-\eta\gamma^{t}\left\Vert \boldsymbol{d}^{t}\right\Vert ^{2}\nonumber \\
 & +\gamma^{t}\left(\bar{f}_{i}^{t}\left(\bar{\boldsymbol{x}}^{t}\right)-\bar{f}_{i}^{t}\left(\boldsymbol{x}^{t}\right)\right)+o\left(\gamma^{t}\right)\nonumber \\
 & \leq f\left(\boldsymbol{x}^{t}\right)-\eta\gamma^{t}\left\Vert \boldsymbol{d}^{t}\right\Vert ^{2}+o\left(\gamma^{t}\right),\forall i=1,...,m\label{eq:fxdecre}
\end{align}
where $o\left(\gamma^{t}\right)$ means that $\lim_{t\rightarrow\infty}o\left(\gamma^{t}\right)/\gamma^{t}=0$.
In (\ref{eq:fxdecre}-a), we used (\ref{eq:ftdbound}) and $\lim_{t\rightarrow\infty}\left\Vert \nabla^{T}f_{i}\left(\boldsymbol{x}^{t}\right)-\nabla^{T}\bar{f}_{i}^{t}\left(\boldsymbol{x}^{t}\right)\right\Vert =0$,
and the last inequality follows from $f_{i}\left(\boldsymbol{x}^{t}\right)\leq f\left(\boldsymbol{x}^{t}\right)$,
$\liminf_{t\rightarrow\infty}f\left(\boldsymbol{x}^{t}\right)-\bar{f}_{i}^{t}\left(\bar{\boldsymbol{x}}^{t}\right)\geq0$,
and $\lim_{t\rightarrow\infty}\left\Vert f_{i}\left(\boldsymbol{x}^{t}\right)-\bar{f}_{i}^{t}\left(\boldsymbol{x}^{t}\right)\right\Vert =0$.
Since (\ref{eq:fxdecre}) holds for all $i=1,...,m$, by choosing
a sufficiently large $t_{\epsilon}$, we have 
\begin{align}
f\left(\boldsymbol{x}^{t+1}\right)-f\left(\boldsymbol{x}^{t}\right) & \leq-\gamma^{t}\overline{\eta}\left\Vert \boldsymbol{d}^{t}\right\Vert ^{2}\nonumber \\
 & \leq-\gamma^{t}\overline{\eta}\epsilon^{'},\forall t\in\mathcal{T}_{\epsilon}^{'}.\label{eq:gapf}
\end{align}
for some $\overline{\eta}>0$. Moreover, from Assumption \ref{asm:convP},
$f\left(\boldsymbol{x}\right)$ is Lipschitz continuous, and thus
\begin{equation}
\left|f\left(\boldsymbol{x}^{t+1}\right)-f\left(\boldsymbol{x}^{t}\right)\right|\leq O(\left\Vert \boldsymbol{x}^{t+1}-\boldsymbol{x}^{t}\right\Vert )\leq O(\gamma^{t})<\epsilon,\label{eq:gapf1}
\end{equation}
$\forall t\geq t_{\epsilon}$, for sufficiently large $t_{\epsilon}$,
where the last inequality follows from $\gamma^{t}\rightarrow0$ as
$t\rightarrow\infty$. From (\ref{eq:gapf}), we know that $f\left(\boldsymbol{x}^{t}\right)$
will be decreased (almost surely) whenever $f\left(\boldsymbol{x}^{t}\right)\geq\epsilon$
and $t\geq t_{\epsilon}$. Therefore, it follows from (\ref{eq:gapf})
and (\ref{eq:gapf1}) that 
\begin{equation}
f\left(\boldsymbol{x}^{t}\right)\leq2\epsilon,\forall t\geq t_{\epsilon}.\label{eq:fxtbound}
\end{equation}
Since (\ref{eq:fxtbound}) is true for any $\epsilon>0$, it follows
that $\limsup_{t\rightarrow\infty}f\left(\boldsymbol{x}^{t}\right)\leq0$.

2. Then we prove that $\lim_{t\rightarrow\infty}\left\Vert \bar{\boldsymbol{x}}^{t}-\boldsymbol{x}^{t}\right\Vert =0,$
w.p.1. 

2.1: We first prove that $\liminf_{t\rightarrow\infty}\left\Vert \bar{\boldsymbol{x}}^{t}-\boldsymbol{x}^{t}\right\Vert =0$
w.p.1.

Note that the feasible problem in (\ref{eq:Pitert-1}) is strictly
convex and thus the solution is uniquely given by $\bar{\boldsymbol{x}}^{t}$.
Therefore, when a feasible update is performed at iteration $t$,
we have $\bar{f}^{t}\left(\bar{\boldsymbol{x}}^{t}\right)\geq0$ and
\begin{align*}
\bar{\boldsymbol{x}}^{t}=\underset{\boldsymbol{x}\in\mathcal{X}}{\text{argmin}\:} & \bar{f}_{0}^{t}\left(\boldsymbol{x}\right)\\
s.t.\: & \bar{f}_{i}^{t}\left(\boldsymbol{x}\right)\leq\bar{f}^{t}\left(\bar{\boldsymbol{x}}^{t}\right),i=1,....,m.
\end{align*}
As a result, $\bar{\boldsymbol{x}}^{t}$ can be expressed in a unified
way as
\begin{align}
\bar{\boldsymbol{x}}^{t}=\underset{\boldsymbol{x}\in\mathcal{X}}{\text{argmin}}\: & \bar{f}_{0}^{t}\left(\boldsymbol{x}\right)\label{eq:Pixbart}\\
s.t.\: & \bar{f}_{i}^{t}\left(\boldsymbol{x}\right)\leq\alpha^{t},i=1,....,m.\nonumber 
\end{align}
where $\alpha^{t}=0$ when an objective update is performed and $\alpha^{t}=\bar{f}^{t}\left(\bar{\boldsymbol{x}}^{t}\right)$
when a feasible update is performed. Since $\lim_{t\rightarrow\infty}\left|\bar{f}^{t}\left(\boldsymbol{x}^{t}\right)-f\left(\boldsymbol{x}^{t}\right)\right|=0$,
$\bar{f}^{t}\left(\bar{\boldsymbol{x}}^{t}\right)\leq\bar{f}^{t}\left(\boldsymbol{x}^{t}\right)$,
and we have proved that $\limsup_{t\rightarrow\infty}f\left(\boldsymbol{x}^{t}\right)\leq0$,
it follows that $\lim_{t\rightarrow\infty}\alpha^{t}=0$. Let $\hat{\boldsymbol{x}}^{t}$
denote the projection of $\boldsymbol{x}^{t}$ on to the feasible
set of Problem (\ref{eq:Pixbart}). Then it follows from $\lim_{t\rightarrow\infty}\alpha^{t}=0$,
$\limsup_{t\rightarrow\infty}\bar{f}^{t}\left(\boldsymbol{x}^{t}\right)=\limsup_{t\rightarrow\infty}f\left(\boldsymbol{x}^{t}\right)\leq0$,
and the strong convexity of $\bar{f}^{t}\left(\boldsymbol{x}^{t}\right)$
that 
\begin{equation}
\lim_{t\rightarrow\infty}\left\Vert \boldsymbol{x}^{t}-\hat{\boldsymbol{x}}^{t}\right\Vert =0.\label{eq:gapxhx}
\end{equation}
From Assumption \ref{asm:convAnew}, $\bar{f}_{0}^{t}\left(\boldsymbol{x}\right)$
is uniformly strongly convex, and thus
\begin{align}
\nabla^{T}\bar{f}_{0}^{t}\left(\boldsymbol{x}^{t}\right)\boldsymbol{d}^{t} & \leq-\eta\left\Vert \boldsymbol{d}^{t}\right\Vert ^{2}+\bar{f}_{0}^{t}\left(\bar{\boldsymbol{x}}^{t}\right)-\bar{f}_{0}^{t}\left(\boldsymbol{x}^{t}\right)\nonumber \\
 & =-\eta\left\Vert \boldsymbol{d}^{t}\right\Vert ^{2}+\bar{f}_{0}^{t}\left(\bar{\boldsymbol{x}}^{t}\right)-\bar{f_{0}}^{t}\left(\hat{\boldsymbol{x}}^{t}\right)\nonumber \\
 & +\bar{f_{0}}^{t}\left(\hat{\boldsymbol{x}}^{t}\right)-\bar{f}_{0}^{t}\left(\boldsymbol{x}^{t}\right)\nonumber \\
 & \leq-\eta\left\Vert \boldsymbol{d}^{t}\right\Vert ^{2}+e\left(t\right),\label{eq:ftdbound-1}
\end{align}
for some $\eta>0$, where $\boldsymbol{d}^{t}=\bar{\boldsymbol{x}}^{t}-\boldsymbol{x}^{t}$,
$\lim_{t\rightarrow\infty}e\left(t\right)=0$, and the last equality
follows from (\ref{eq:gapxhx}). From Assumption \ref{asm:convP},
the gradient of $f_{0}\left(\boldsymbol{x}\right)$ is Lipschitz continuous,
and thus there exists $L_{0}>0$ such that
\begin{align*}
f_{0}\left(\boldsymbol{x}^{t+1}\right) & \leq f_{0}\left(\boldsymbol{x}^{t}\right)+\gamma^{t}\nabla^{T}f_{0}\left(\boldsymbol{x}^{t}\right)\boldsymbol{d}^{t}+L_{0}\left(\gamma^{t}\right)^{2}\left\Vert \boldsymbol{d}^{t}\right\Vert ^{2}\\
 & =f_{0}\left(\boldsymbol{x}^{t}\right)+L_{0}\left(\gamma^{t}\right)^{2}\left\Vert \boldsymbol{d}^{t}\right\Vert ^{2}\\
 & +\gamma^{t}\left(\nabla^{T}f_{0}\left(\boldsymbol{x}^{t}\right)-\nabla^{T}\bar{f_{0}}^{t}\left(\boldsymbol{x}^{t}\right)+\nabla^{T}\bar{f_{0}}^{t}\left(\boldsymbol{x}^{t}\right)\right)\boldsymbol{d}^{t}\\
 & \leq f_{0}\left(\boldsymbol{x}^{t}\right)-\gamma^{t}\eta\left\Vert \boldsymbol{d}^{t}\right\Vert ^{2}+o\left(\gamma^{t}\right)
\end{align*}
where in the last inequality, we used (\ref{eq:ftdbound-1}) and $\lim_{t\rightarrow\infty}\left\Vert \nabla^{T}f_{0}\left(\boldsymbol{x}^{t}\right)-\nabla^{T}\bar{f}_{0}^{t}\left(\boldsymbol{x}^{t}\right)\right\Vert =0$.
Let us show by contradiction that w.p.1. $\liminf_{t\rightarrow\infty}\left\Vert \bar{\boldsymbol{x}}^{t}-\boldsymbol{x}^{t}\right\Vert =0$.
Suppose $\liminf_{t\rightarrow\infty}\left\Vert \bar{\boldsymbol{x}}^{t}-\boldsymbol{x}^{t}\right\Vert \geq\chi>0$
with a positive probability. Then we can find a realization such that
$\left\Vert \boldsymbol{d}^{t}\right\Vert \geq\chi$ at the same time
for all $t$. We focus next on such a realization. By choosing a sufficiently
large $t_{0}$, there exists $\overline{\eta}>0$ such that
\begin{align}
f_{0}\left(\boldsymbol{x}^{t+1}\right)-f_{0}\left(\boldsymbol{x}^{t}\right) & \leq-\gamma^{t}\overline{\eta}\left\Vert \boldsymbol{d}^{t}\right\Vert ^{2},\forall t\geq t_{0}.\label{eq:gapf0xt}
\end{align}
It follows from (\ref{eq:gapf0xt}) that 
\[
f_{0}\left(\boldsymbol{x}^{t}\right)-f_{0}\left(\boldsymbol{x}^{t_{0}}\right)\leq-\overline{\eta}\chi^{2}\sum_{j=t_{0}}^{t}\gamma^{j},
\]
which, in view of $\sum_{j=t_{0}}^{\infty}\gamma^{j}=\infty$, contradicts
the boundedness of $\left\{ f_{0}\left(\boldsymbol{x}^{t}\right)\right\} $.
Therefore it must be $\liminf_{t\rightarrow\infty}\left\Vert \bar{\boldsymbol{x}}^{t}-\boldsymbol{x}^{t}\right\Vert =0$
w.p.1.

2.2: Then we prove that $\limsup_{t\rightarrow\infty}\left\Vert \bar{\boldsymbol{x}}^{t}-\boldsymbol{x}^{t}\right\Vert =0$
w.p.1.

We first prove a useful lemma.
\begin{lem}
\label{lem:gapxbar}There exists a constant $\hat{L}>0$ such that
\[
\left\Vert \bar{\boldsymbol{x}}^{t_{1}}-\bar{\boldsymbol{x}}^{t_{2}}\right\Vert \leq\hat{L}\left\Vert \boldsymbol{x}^{t_{1}}-\boldsymbol{x}^{t_{2}}\right\Vert +e\left(t_{1},t_{2}\right),
\]
where $\lim_{t_{1},t_{2}\rightarrow\infty}e\left(t_{1},t_{2}\right)=0$.
\end{lem}

\begin{IEEEproof}
From Assumption \ref{asm:convAnew}-2 and \ref{asm:convP}-2, we have
\begin{equation}
\left|\bar{f}_{i}^{t_{1}}\left(\boldsymbol{x}\right)-\bar{f}_{i}^{t_{2}}\left(\boldsymbol{x}\right)\right|\leq B\left\Vert \boldsymbol{x}^{t_{1}}-\boldsymbol{x}^{t_{2}}\right\Vert +e^{'}\left(t_{1},t_{2}\right),\label{eq:gapft1t2}
\end{equation}
for all $\boldsymbol{x}\in\mathcal{X}$ and $i=0,1,...,m$, where
$\lim_{t_{1},t_{2}\rightarrow\infty}e^{'}\left(t_{1},t_{2}\right)=0$.
Then it follows from (\ref{eq:gapft1t2}) and (\ref{eq:Pixbart}),
and the Lipschitz continuity and strong convexity of $\bar{f}_{i}^{t}\left(\boldsymbol{x}\right),\forall i$
that
\begin{equation}
\left\Vert \bar{\boldsymbol{x}}^{t_{1}}-\bar{\boldsymbol{x}}^{t_{2}}\right\Vert \leq B_{1}B\left\Vert \boldsymbol{x}^{t_{1}}-\boldsymbol{x}^{t_{2}}\right\Vert +B_{1}e^{'}\left(t_{1},t_{2}\right)+B_{2}\alpha^{t},\label{eq:xt1t2}
\end{equation}
for some constant $B_{1},B_{2}>0$. This is because for the strictly
convex problem in (\ref{eq:Pixbart}) with Lipschitz continuous and
strongly convex objective/constraint functions, when the objective
and constraint functions in (\ref{eq:Pixbart}) are changed by some
amount $e_{i}\left(\boldsymbol{x}\right),i=0,1,...,m$, the optimal
solution $\bar{\boldsymbol{x}}^{t}$ will be changed by the same order,
i.e., the change is within the range $\pm O\left(max_{i}\left|e_{i}\left(\boldsymbol{x}\right)\right|\right)$.
Finally, Lemma \ref{lem:gapxbar} follows from (\ref{eq:xt1t2}) immediately.
\end{IEEEproof}

Using Lemma \ref{lem:gapxbar} and following the same analysis as
that in (\cite{Yang_TSP2016_SSCA}, Proof of Theorem 1), it can be
shown that \textbf{$\limsup_{t\rightarrow\infty}\left\Vert \bar{\boldsymbol{x}}^{t}-\boldsymbol{x}^{t}\right\Vert =0$
}w.p.1.

This completes the proof.



\begin{thebibliography}{10}
\providecommand{\url}[1]{#1}
\csname url@samestyle\endcsname
\providecommand{\newblock}{\relax}
\providecommand{\bibinfo}[2]{#2}
\providecommand{\BIBentrySTDinterwordspacing}{\spaceskip=0pt\relax}
\providecommand{\BIBentryALTinterwordstretchfactor}{4}
\providecommand{\BIBentryALTinterwordspacing}{\spaceskip=\fontdimen2\font plus
\BIBentryALTinterwordstretchfactor\fontdimen3\font minus
  \fontdimen4\font\relax}
\providecommand{\BIBforeignlanguage}[2]{{%
\expandafter\ifx\csname l@#1\endcsname\relax
\typeout{** WARNING: IEEEtran.bst: No hyphenation pattern has been}%
\typeout{** loaded for the language `#1'. Using the pattern for}%
\typeout{** the default language instead.}%
\else
\language=\csname l@#1\endcsname
\fi
#2}}
\providecommand{\BIBdecl}{\relax}
\BIBdecl

\bibitem{Spall_Wiley03_SO}
J.~C. Spall, \emph{Introduction to Stochastic Search and Optimization:
  Estimation, Simulation and Control}.\hskip 1em plus 0.5em minus 0.4em\relax
  Hoboken, NJ: Wiley, 2003.

\bibitem{Bertsekas_SIAM2000_SG}
D.~P. Bertsekas and J.~N. Tsitsiklis, ``Gradient convergence in gradient
  methods with errors,'' \emph{SIAM J. Optim.}, vol.~10, no.~3, pp. 627--642,
  2000.

\bibitem{Polyak_SIAM1992_SO}
B.~T. Polyak and A.~B. Juditsky, ``Acceleration of stochastic approximation by
  averaging,'' \emph{SIAM Journal on Control and Optimization}, vol.~30, no.~4,
  pp. 838--855, 1992.

\bibitem{Ram_Asilomar07_ISG}
S.~S. Ram, A.~Nedic, and V.~V. Veeravalli, ``Stochastic incremental gradient
  descent for estimation in sensor networks,'' in \emph{2007 Conference Record
  of the Forty-First Asilomar Conference on Signals, Systems and Computers},
  Nov. 2007, pp. 582--586.

\bibitem{NIPS2013_4937}
R.~Johnson and T.~Zhang, ``Accelerating stochastic gradient descent using
  predictive variance reduction,'' in \emph{Advances in Neural Information
  Processing Systems 26}, C.~J.~C. Burges, L.~Bottou, M.~Welling,
  Z.~Ghahramani, and K.~Q. Weinberger, Eds., 2013, pp. 315--323.

\bibitem{NIPS2014_5258}
A.~Defazio, F.~Bach, and S.~Lacoste-Julien, ``{SAGA}: A fast incremental
  gradient method with support for non-strongly convex composite objectives,''
  in \emph{Advances in Neural Information Processing Systems 27}, 2014, pp.
  1646--1654.

\bibitem{Ermoliev_Cybern1972_SG}
Y.~Ermoliev, ``On the method of generalized stochastic gradients and
  quasi-fejer sequences,'' \emph{Cybern.}, vol.~5, no.~2, pp. 208--220, 1972.

\bibitem{Yousefian_Automatica2012_SG}
F.~Yousefian, A.~Nedic, and U.~V. Shanbhag, ``On stochastic gradient and
  subgradient methods with adaptive steplength sequences,'' \emph{Automatica},
  vol.~48, no.~1, pp. 56--67, 2012.

\bibitem{Ruszczyski_MP1980_SAG}
A.~Ruszczynski, ``Feasible direction methods for stochastic programming
  problems,'' \emph{Math. Programm.}, vol.~19, no.~1, pp. 220--229, Dec. 1980.

\bibitem{Bach_2014AAS_SAG}
F.~Bach, ``Adaptivity of averaged stochastic gradient descent to local strong
  convexity for logistic regression,'' \emph{J. Mach. Learn. Res.}, vol.~15,
  no.~1, pp. 595--627, Jan 2014.

\bibitem{Gupal_Cyber1972_SAG}
A.~M. Gupal and L.~G. Bazhenov, ``Stochastic analog of the conjugant gradient
  method,'' \emph{Cybernetics}, vol.~8, no.~1, pp. 138--140, 1972.

\bibitem{Yin_SSR_1994_SGA}
G.~Yin and K.~Yin, ``Asymptotically optimal rate of convergence of smoothed
  stochastic recursive algorithms,'' \emph{Stochastics and Stochastic Reports},
  vol.~47, no. 1-2, pp. 21--46, 1994.

\bibitem{Yin1995}
G.~Yin, \emph{Adaptive Filtering with Averaging}.\hskip 1em plus 0.5em minus
  0.4em\relax New York, NY: Springer New York, 1995, pp. 375--396.

\bibitem{Sun_TSP2017_MM}
Y.~Sun, P.~Babu, and D.~P. Palomar, ``Majorization-minimization algorithms in
  signal processing, communications, and machine learning,'' \emph{IEEE
  Transactions on Signal Processing}, vol.~65, no.~3, pp. 794--816, Feb 2017.

\bibitem{Bertsekas_TR2010_proximal}
D.~P. Bertsekas, ``Incremental gradient, subgradient, and proximal methods for
  convex optimization: A survey,'' \emph{MIT, Cambridge, MA, LIDS Tech. Rep.},
  2010.

\bibitem{Olivier_JRS2009_EM}
O.~Cappe and E.~Moulines, ``On-line expectation-maximization algorithm for
  latent data models,'' \emph{Journal of the Royal Statistical Society. Series
  B (Statistical Methodology)}, vol.~71, no.~3, pp. 593--613, 2009.

\bibitem{Stoica_spm2004_CyclicMIN}
P.~Stoica and Y.~Selen, ``Cyclic minimizers, majorization techniques, and the
  expectation-maximization algorithm: a refresher,'' \emph{IEEE Signal
  Processing Magazine}, vol.~21, no.~1, pp. 112--114, Jan 2004.

\bibitem{Wainwright_FTML2008_VBI}
M.~J. Wainwright and M.~I. Jordan, ``Graphical models, exponential families,
  and variational inference,'' \emph{Found. Trends Mach. Learn.}, vol.~1, no.
  1-2, pp. 1--305, Jan 2008.

\bibitem{NIPS2013_5129_StochasticMM}
J.~Mairal, ``Stochastic majorization-minimization algorithms for large-scale
  optimization,'' in \emph{Advances in Neural Information Processing Systems
  26}, 2013, pp. 2283--2291.

\bibitem{Chouzenoux_TSP17_StochasticMM}
E.~Chouzenoux and J.~C. Pesquet, ``A stochastic majorize-minimize subspace
  algorithm for online penalized least squares estimation,'' \emph{IEEE
  Transactions on Signal Processing}, vol.~65, no.~18, pp. 4770--4783, Sept
  2017.

\bibitem{Scutari_TSP14_SCA}
G.~Scutari, F.~Facchinei, P.~Song, D.~P. Palomar, and J.~S. Pang,
  ``Decomposition by partial linearization: Parallel optimization of
  multi-agent systems,'' \emph{IEEE Trans. Signal Processing}, vol.~62, no.~3,
  pp. 641--656, Feb 2014.

\bibitem{Yang_TSP2016_SSCA}
Y.~Yang, G.~Scutari, D.~P. Palomar, and M.~Pesavento, ``A parallel
  decomposition method for nonconvex stochastic multi-agent optimization
  problems,'' \emph{IEEE Trans. Signal Processing}, vol.~64, no.~11, pp.
  2949--2964, June 2016.

\bibitem{Nemirovski_SIAM2006_CCp}
A.~Nemirovski and A.~Shapiro, ``Convex approximations of chance constrained
  programs,'' \emph{SIAM J. Optim.}, vol.~17, no.~4, pp. 969--996, 2006.

\bibitem{Ding_TSP09_MIMOimpCSI}
M.~Ding and S.~D. Blostein, ``{MIMO} minimum total {MSE} transceiver design
  with imperfect {CSI} at both ends,'' \emph{IEEE Trans. Signal Processing},
  vol.~57, no.~3, pp. 1141--1150, March 2009.

\bibitem{Wang_ICASSP11_Bernstein-type}
K.-Y. Wang, T.-H. Chang, W.-K. Ma, A.-C. So, and C.-Y. Chi, ``Probabilistic
  {SINR} constrained robust transmit beamforming: A {Bernstein-type} inequality
  based conservative approach,'' in \emph{in Proc. IEEE ICASSP 2011}, May.
  2011, pp. 3080--3083.

\bibitem{Liu_TSP14_RFprecoding}
A.~Liu and V.~K.~N. Lau, ``Phase only {RF} precoding for massive {MIMO} systems
  with limited {RF} chains,'' \emph{IEEE Trans. Signal Processing}, vol.~62,
  no.~17, pp. 4505--4515, Sept. 2014.

\bibitem{Liu_TSP2016_CSImassive}
------, ``Impact of {CSI} knowledge on the codebook-based hybrid beamforming in
  massive {MIMO},'' \emph{IEEE Transactions on Signal Processing}, vol.~64,
  no.~24, pp. 6545--6556, Dec 2016.

\bibitem{Zhang_TSP05_RFshifter}
X.~Zhang, A.~Molisch, and S.-Y. Kung, ``Variable-phase-shift-based
  {RF-baseband} codesign for {MIMO} antenna selection,'' \emph{IEEE Trans.
  Signal Processing}, vol.~53, no.~11, pp. 4091--4103, Nov. 2005.

\bibitem{Meisam_thesis14_BSUM}
M.~Razaviyayn, ``Successive convex approximation: Analysis and applications,''
  Ph.D. dissertation, University of Minnesota, 2014.

\bibitem{SPlecbook}
A.~Shapiro, D.~Dentcheva, and A.~Ruszczynski, \emph{Lectures on Stochastic
  Programming: Modeling and Theory}, ser. MPS-SIAM Series on
  Optimization.\hskip 1em plus 0.5em minus 0.4em\relax SIAM-Society for
  Industrial and Applied Mathematics, September 2009.

\bibitem{Mahdavi2013Online}
M.~Mahdavi, T.~Yang, and R.~Jin, ``Online stochastic optimization with multiple
  objectives,'' \emph{Advances in Neural Information Processing Systems}, 2013.

\bibitem{cvx}
M.~Grant and S.~Boyd, ``{CVX}: Matlab software for disciplined convex
  programming, version 2.1,'' \url{http://cvxr.com/cvx}, Mar. 2014.

\bibitem{Ruszczyski_MP80_SPthem}
A.~Ruszczynski, ``Feasible direction methods for stochastic programming
  problems,'' \emph{Math. Programm.}, vol.~19, no.~1, pp. 220--229, Dec. 1980.

\bibitem{Dunford_Inpub58_LO}
N.~Dunford and J.~T. Schwartz, \emph{Linear Operators. Part 1: General
  Theory}.\hskip 1em plus 0.5em minus 0.4em\relax Interscience Publ. New York,
  1958.

\end{thebibliography}
\end{document}